\documentclass{egpubl}
\usepackage{eurovis2023}

\SpecialIssuePaper         

\CGFStandardLicense

\usepackage[T1]{fontenc}
\usepackage{dfadobe}  

\usepackage{cite}  
\BibtexOrBiblatex
\electronicVersion
\PrintedOrElectronic
\ifpdf \usepackage[pdftex]{graphicx} \pdfcompresslevel=9
\else \usepackage[dvips]{graphicx} \fi

\usepackage{egweblnk}

\usepackage{xcolor}
\usepackage{subcaption}
\newcommand{\rev}[1]{{\color{black}{#1}}} 

\title[Illustrative Motion Smoothing]%
      {Illustrative Motion Smoothing for \\Attention Guidance in Dynamic Visualizations}

\author[J. Eschner et al.]
{\parbox{\textwidth}{\centering 
        Johannes Eschner$^1$\orcid{0009-0001-6784-8503}
        Peter Mindek$^2$\orcid{0000-0002-9434-5952} 
        Manuela Waldner$^1$\orcid{0000-0003-1387-5132}  
        }
         \\
 {\parbox{\textwidth}{\centering $^1$TU Wien, Austria  $^2$Nanographics GmbH, Austria\\
        } 
 }
 }

\begin{document}

\teaser{
    \centering
        \subfloat[]{
             \includegraphics[width=0.325\textwidth]{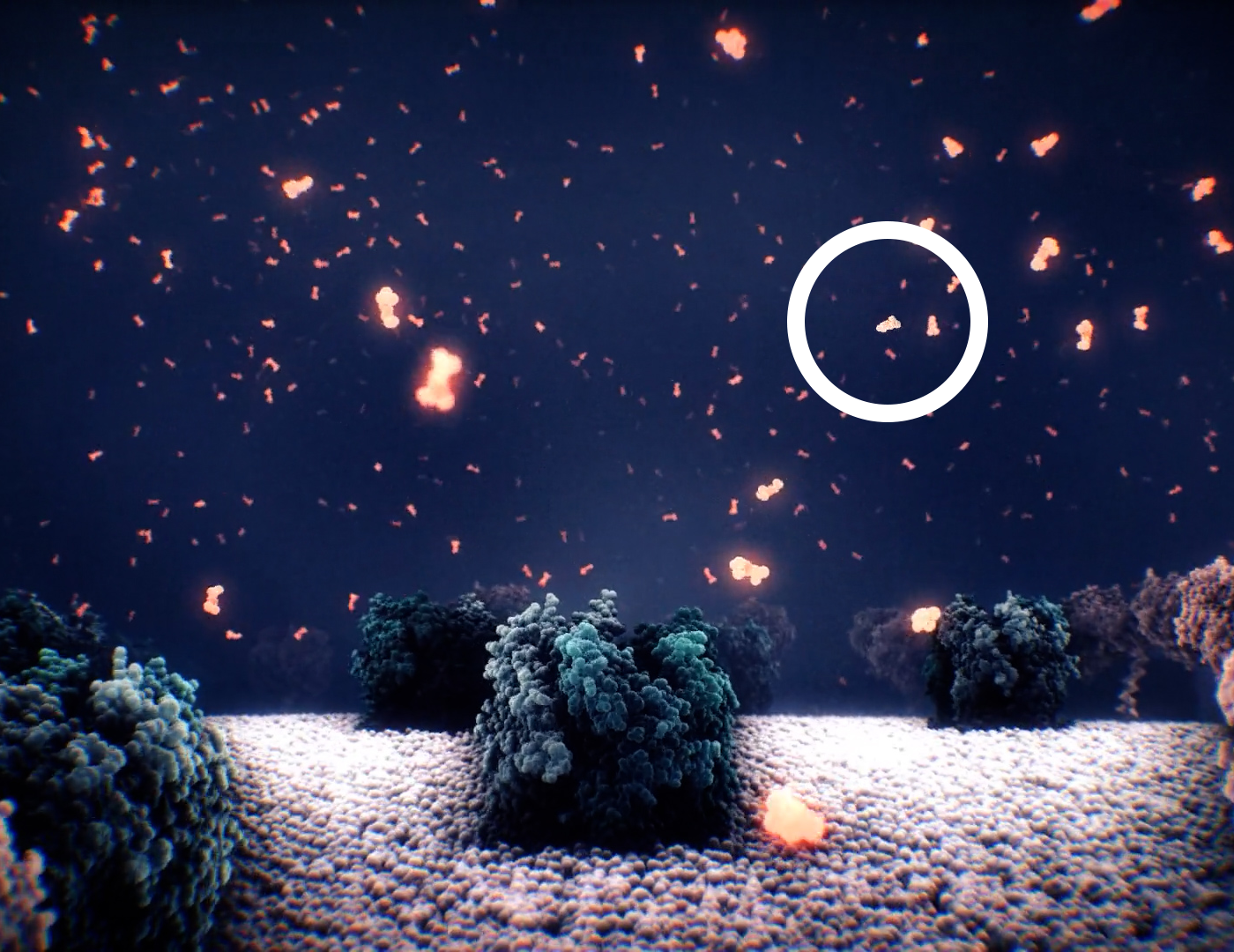}
             \label{fig:teaser_tl0}
        } ~
         \hspace{-1.1em}
        \subfloat[]{
             \includegraphics[width=0.325\textwidth]{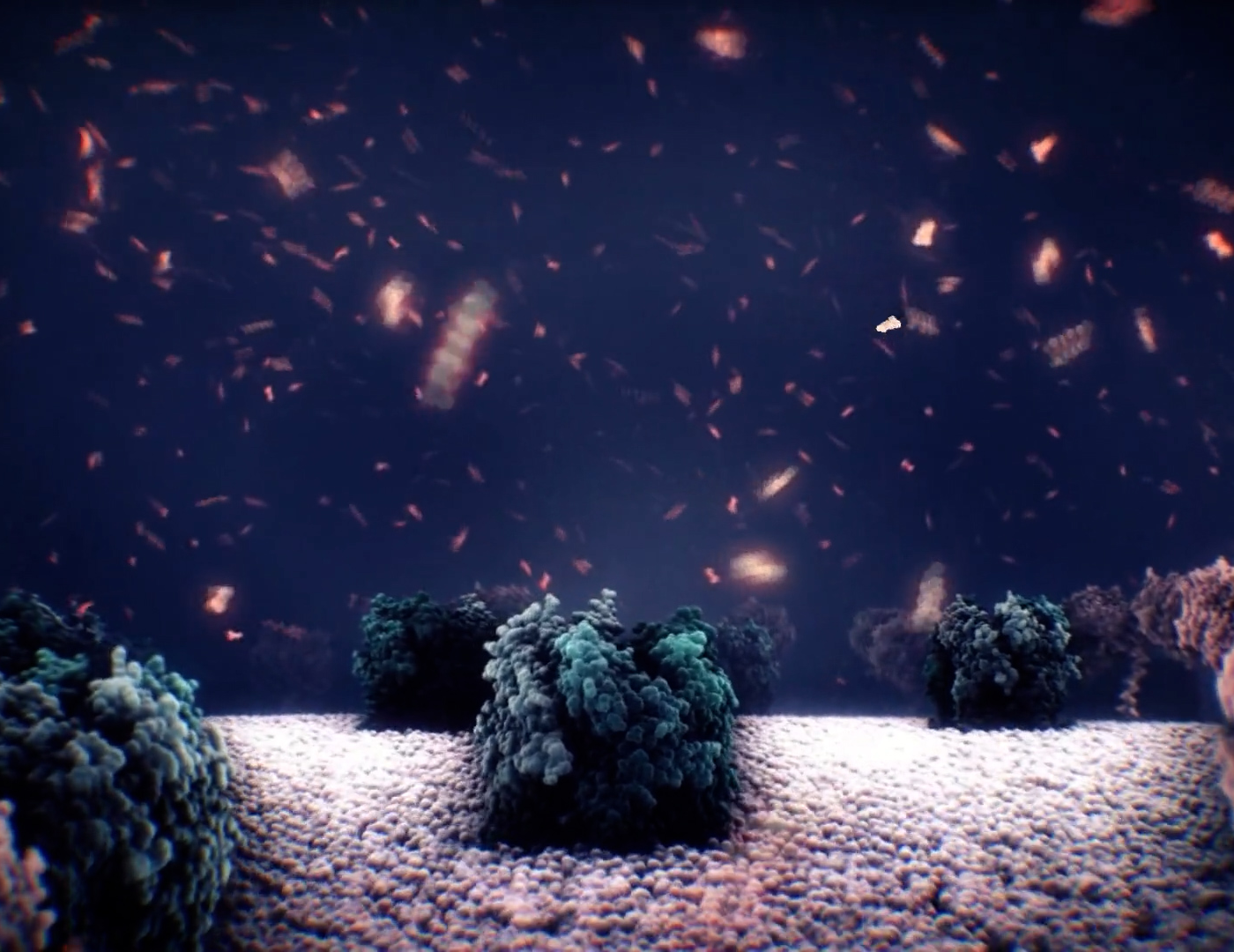}
             \label{fig:teaser_tl2}
        } ~
        \hspace{-1.1em}
        \subfloat[]{
             \includegraphics[width=0.325\textwidth]{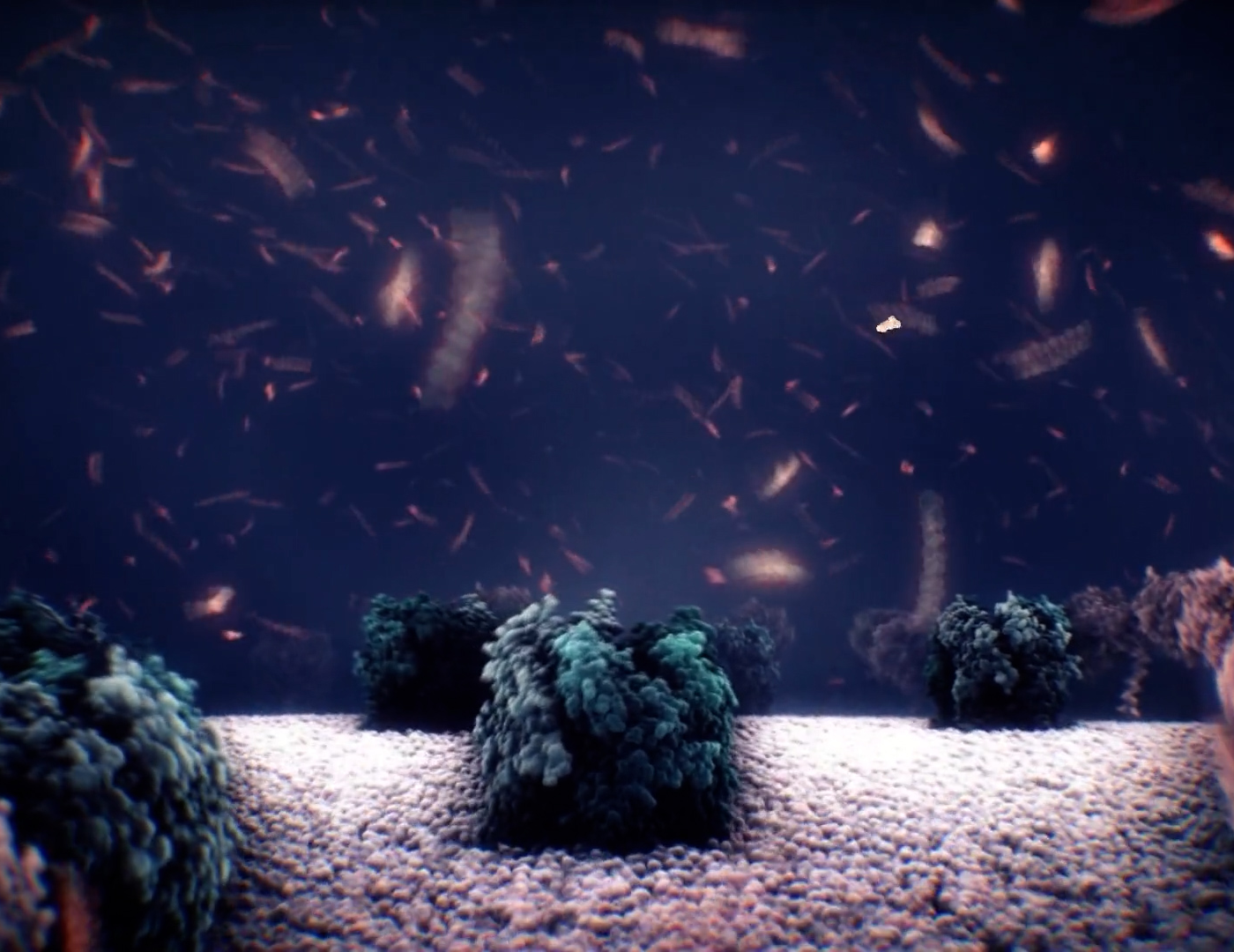}
             \label{fig:teaser_tl4}
        } 
    \caption{Snapshot of a single frame of a molecular animation without any illustrative motion smoothing (a), with moderate visual motion smoothing applied to context elements (b), and a high level of visual motion smoothing (c). The focus element is marked by a circle in (a) for illustration purposes. Moderate visual motion smoothing (b) was found to be a good trade-off between story comprehensiveness and obtrusiveness in our study. Please also refer to the supplementary video. }
    \label{fig:teaser}
}

\maketitle
\begin{abstract}
   3D animations are an effective method to learn about complex dynamic phenomena, such as mesoscale biological processes.  The animators' goals are to convey a sense of the scene's overall complexity while, at the same time, visually guiding the user through a story of subsequent events embedded in the chaotic environment. Animators use a variety of visual emphasis techniques to guide the observers' attention through the story, such as highlighting, halos -- or by manipulating motion parameters of the scene. In this paper, we investigate the effect of smoothing the motion of contextual scene elements to attract attention to focus elements of the story exhibiting high-frequency motion. We conducted a crowdsourced study with \rev{108} participants observing short animations with two illustrative motion smoothing strategies: geometric smoothing \rev{through noise reduction of contextual motion trajectories} and visual smoothing through motion blur of \rev{context} items. We investigated the observers' ability to follow the story as well as the effect of the techniques on speed perception in a molecular scene. Our results show that moderate motion blur significantly improves users' ability to follow the story. Geometric motion smoothing \rev{is less effective} but increases the visual appeal of the animation. However, both techniques also slow down the perceived speed of the animation. We discuss the implications of these results and derive design guidelines for animators of complex dynamic visualizations. 


\end{abstract}
\section{Introduction}
\label{sec:introduction}

Learning about dynamic processes can be facilitated by animation, which has been found to be more effective than a series of static images when done well~\cite{tversky2002animation}. For example, molecular visualization frequently uses animations to depict complex biological processes~\cite{kozlikova2017visualization}. In the field of biology, studies have shown that students retain more information for a longer period of time after viewing an animation instead of static graphics~\cite{o2007value} and that students had the best post-test scores -- and thus understanding of the depicted biological processes -- after watching a very detailed animation compared to more abstracted ones~\cite{jenkinson2012visualizing}. 

Visualization of dynamic biological processes is particularly challenging because the phenomena to be conveyed often consist of a series of fast and chaotic movements \rev{ -- the so-called \emph{Brownian motion} --} in a crowded environment. \rev{This seemingly erratic movement by individual molecules is caused by constant collisions with other molecules.} Therefore, biological animations \rev{easily become} overwhelming or cluttered. That means they are so packed with visual information that it is getting difficult to reliably attract the viewer's attention to the main actors of the story~\cite{rosenholtz2007measuring}. If too cluttered, an animation may even be harmful for learning, especially if the content is so complex that it is unclear where the attention should be directed~\cite{ruiz2009computer}. Animators are therefore facing the challenge of carefully balancing two competing requirements for creating insightful biological animations: they want to communicate the complexity of the scene and, at the same time, visually guide the user through the story of subsequent events embedded in this chaotic environment. 

Visualization designers and animators use a variety of techniques to declutter complex dynamic visualizations, such as lowlighting of contextual parts of the animation~\cite{de2007attention}, visual abstraction~\rev{\cite{byvska2019analysis,lawonn2014line,garrison2021exploration}}, or sparsification and carefully chosen camera angles~\cite{iwasa2007visualizing}. \rev{Popular educational videos in the biological field often use a combination of selective camera angles, lowlighting, depth-of-field, and motion blur~\cite{berry2009malaria,berry2018dna}.} While design recommendations~\cite{tversky2002animation,ruiz2009computer} and experience reports~\cite{iwasa2007visualizing} exist, empirical evidence on attention guidance in dynamic scenes is scarce \rev{such} that animators often work based on their intuition. 
Indeed, in their survey of 25 years of animation, Chevalier et al.~\cite{chevalier2016animations} note that the influence of different factors of the animation, such as the speed or the number of objects possibly tracked, is still not sufficiently investigated.

In this work, we make a step towards closing this gap by systematically investigating the effect of \emph{illustrative motion smoothing} on the story comprehensiveness, speed perception, and subjective visual appeal of a dynamic biological visualization. Motion smoothing declutters the scene by removing high-frequency movement. We are particularly interested \rev{in whether} selectively applying motion smoothing effects is sufficient to effectively guide the users' attention without adding artificial attention cues, such as halos, to the scene. We propose and investigate two motion smoothing strategies: 1) \emph{geometric} smoothing that removes high-frequency jitter of the scene elements' motion trajectories, and 2) \emph{visual} smoothing by adding motion blur. Our proposed illustrative motion smoothing applies these smoothing effects to contextual scene elements only while retaining the unmodified appearance of focus elements. 

Through a crowdsourced study investigating a dynamic molecular visualization, we aim to answer the following research questions: 

\begin{description}
    \item[Q1 (comprehensiveness):] Is illustrative motion smoothing sufficient to visually guide users through the main story in a cluttered dynamic visualization without additional attention cues?  
    \item[Q2 (perception):] How does illustrative motion smoothing affect the perceived speed of the dynamic visualization?  
    \item[Q3 (obtrusiveness):] Does illustrative motion smoothing affect the visual quality of the visualized scene? 
\end{description}

In summary, our contributions are the following:
\rev{
\begin{enumerate}
    \item We introduce a new attention guidance technique, which selectively applies motion smoothing to contextual scene elements in crowded dynamic visualizations.
    \item We present the results of a crowdsourced study evaluating the effectiveness of illustrative motion smoothing, its impact on speed perception, and its level of obtrusiveness.
    \item We formulate design guidelines for effective illustrative motion smoothing in dynamic visualizations.
\end{enumerate}
}

\section{Related Work}

We first outline how previous work has used illustrative techniques to effectively convey dynamic phenomena in static and dynamic visualizations. Afterwards, we discuss the background of motion perception in visualizations. 

\subsection{Visualization of Dynamic Processes}

Static visualizations depicting dynamic processes often use illustrative techniques to indicate motion, such as speedlines, flow ribbons, opacity modulation, or strobe silhouettes~\cite{joshi2005illustration}. Others explicitly illustrate ``causal chains'' by analyzing the motion of scene elements, such as mechanical assemblies or motion-captured body parts, and sequentially highlight key elements together with arrows~\cite{mitra2010illustrating,bouvier2007motion}. Short video sequences can be visually summarized in single images using stroboscopic motion illustration~\cite{agarwala2004interactive} and visualizations of extracted trajectories~\cite{meghdadi2013interactive}. Stroboscopic motion illustration is conceptually similar to motion blur \rev{as it merges multiple static poses of the animation into a single image to suggest movement}. \rev{Similarly, for} static images, motion blur generates the impression that objects are moving~\cite{chen1996image}. To the best of our knowledge, stroboscopic motion illustration has been demonstrated only on sparse scenes, such as footage from a static surveillance camera. In this work, we are interested in cluttered scenes capturing complex processes. For these cases, animations seem to be more effective~\cite{tversky2002animation,o2007value,jenkinson2012visualizing}. Therefore, we focus on dynamic visualizations in this work.  

Particularly in the field of molecular visualization, designers and researchers have experimented with a variety of illustrative visualization techniques. For example, the molecular viewer ePMV~\cite{johnson2011epmv} supports motion blur both for static images and for slow animations \emph{``to imply dynamics''}. In contrast to global motion blur, we apply motion blur selectively to context elements to make focus elements stand out, similarly to the semantic-depth-of-field effect for static visualizations~\cite{kosara2002focus+}. Le Muzic et al.~\cite{le2015illustrative} performed geometric trajectory smoothing of selected focus molecules in fast-forward animations to allow users to follow a chain of reactions. In our work, we propose a similar geometric motion smoothing effect but reverse it: we do not manipulate the movement of focus elements but rather reduce high-frequency motion of contextual scene elements to make focus elements visually stand out due to their more chaotic movement. 

Others manipulate the rendered visualization by slowing down or speeding up the entire animation -- either depending on a global ~\cite{byvska2019analysis} or a spatial degree of interest~\cite{solteszova2020memento}.
Using such global ``time-stretching'' and ``time compression'' is a common strategy of animators~\cite{jenkinson2017role}. 
In virtual reality, researchers have proposed slow motion effects, \rev{similar to effects} shown in the well-known \emph{Matrix} movie~\cite{rietzler2017matrix,lee2019simulating}. During studies of these systems, users explicitly asked for motion blur effects to emphasize the slow motion episodes. In our work, we apply selective geometric motion smoothing on contextual scene elements and explore the interaction between this geometric smoothing and motion blur.

There is surprisingly little work on attention guidance in dynamic scenes, which does not slow down the entire animation. 
One strategy is to suppress the lightness and contrast of contextual parts of the animation~\cite{de2007attention}. \rev{Others use motion-independent depth-of-field effects for animated pathlines~\cite{lawonn2014adaptive}.} 
Researchers have also investigated attention guidance by adding highly salient attention cues, such as flicker~\cite{waldner2014attractive}, or by using an additional modality to the visual content, such as audio~\cite{rothe2017diegetic,xie2019coordinating}. 
In the field of mixed reality, where users are facing dynamic imagery by constantly changing their view, researchers have proposed attention guidance through luminance changes~\cite{booth2013guiding}, by target saliency manipulation of the image~\cite{sutton2022look}, or by adding an artificially moving light cone~\cite{rothe2017diegetic}. In contrast to these works, we assume that we have a highly complex scene, \rev{whose inherent motion} we aim to selectively declutter to guide the user's attention to the remaining high-frequency elements in the scene. \rev{To the best of our knowledge, we are the first to systematically investigate such selective motion smoothing in dynamic scenes.}

\subsection{Motion Perception in Dynamic Visualizations}

Human observers are very sensitive to motion. 
Studies have shown that in computational neurobiological models of attention (as proposed, for example, by Itti et al.~\cite{itti2003realistic}), flicker and motion detectors are the strongest predictors of human attention in dynamic scenes~\cite{itti2005quantifying,mital2011clustering}. It can therefore be expected that selectively modifying motion parameters in the scene can have considerable effects on scene perception: 

Animation speed can have a strong influence on what users may learn from an animation. Fischer et al.~\cite{fischer2008effects} showed that users focus more on the functional aspects when viewing an animation of a pendulum clock at a higher speed. Selectively smoothing the trajectory of focus elements allows users to better follow the story but also makes the scene appear slower as a whole~\cite{le2015illustrative}. 

The human visual system (HSV) integrates the images captured on the retina for around 120 milliseconds, which can lead to a smeared or elongated perception of quickly moving objects~\cite{burr1980motion}. For static images, such as photographs or illustrations, motion blur therefore can generate the impression that objects are moving~\cite{chen1996image}. In a dynamic scene containing a single rotating sphere, Navarro et al.~\cite{navarro2011perceptual} investigated the effect of shutter times, antialiasing, and texture detail at different movement speeds on perceived visual rendering quality. Their results indicate that too long shutter speeds lead to unrealistically excessive motion blur. In more complex dynamic scenes, objects that are actively tracked by the observer are expected to appear sharper than non-tracked distractors~\cite{stengel2014temporal}. Our illustrative motion smoothing approach takes these considerations into account and leaves the focus elements unsmoothed. 

Tang et al.~\cite{tang2013teleoperated} evaluated the impact of artificial motion blur on speed perception for teleoperated vehicles. They showed that users were significantly better to judge their driving speed in the presence of motion blur. In contrast, adding motion blur to a 3D racing game did not improve the players' subjective speed impression or gaming experience~\cite{sharan2013simulated}. Holm et al.~\cite{holm2016increasing} found that a larger field of view increases the perceived speed in racing games, but strong motion blur can even decrease the perceived speed. In virtual reality scenes, motion or depth-of-field blur did not have any influence on users' distance and speed estimations~\cite{langbehn2016visual}. In summary, these studies provide mixed results about the impact of motion blur on subjective speed perception. However, the visualizations we want to address differ as we have a static observer and a highly dynamic scene.   

Strictly controlled psychophysics experiments have shown that extending the length of a motion smear following an object leads to speed overestimation~\cite{vaziri2008apparent}. 
In the context of flow visualization, Birkeland et al.~\cite{birkeland2014perceptually} evaluated the perception of relative speed for animated flow fields. They varied the speed, contrast, direction of motion, and also added comet tails. The comet tails could slightly improve the users' speed estimations. 
In contrast to these works, dynamic molecular visualization consists of merely undirected, random motion. In the context of surveillance video analysis, Höferlin et al.~\cite{hoferlin2012evaluation} compared four fast-forward video visualization techniques \rev{to summarize long video sequences}: frame skipping, temporal blending to generate motion blur effects, adding object trails, and showing predictive motion cues. In their study, temporal blending led to subjective information overload, which caused users to miss search targets. Also, the ability to perceive motion was considered low by the participants. However, surveillance videos only sparsely contain moving objects, while dynamic molecular scenes are much more crowded. \rev{Our work, therefore, contributes new insights for attention guidance in highly complex dynamic visualizations.}

\begin{figure*}
     \centering
     \begin{subfigure}[]{\linewidth}
         \includegraphics[width=\textwidth]{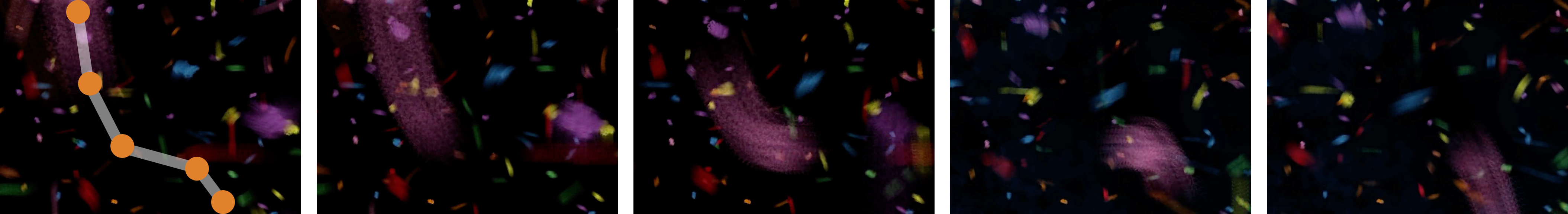}
         \caption{Baseline: Non-smoothed trajectories.}
         \label{fig:smoothness0}
     \end{subfigure}
     \hfill
     \begin{subfigure}[]{\linewidth}
         \includegraphics[width=\textwidth]{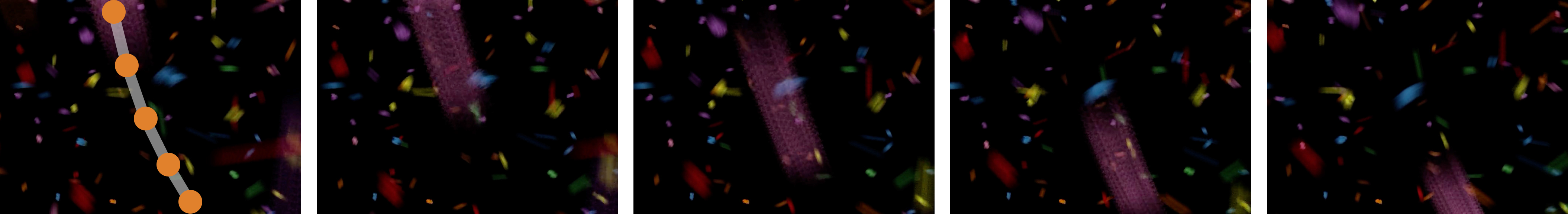}
         \caption{Geometric motion smoothing with smoothed trajectories.}
         \label{fig:smoothness1}
     \end{subfigure}
     
        \caption{A short sequence of frames of a scene with extensive visual motion smoothing (trail length 4), without (a) and with (b) geometric motion smoothing. In the setting without geometric motion smoothing (with $\tau=0$ in Equation~\ref{eq:smoothMotion}), Brownian motion leads to extensive jitter of the pink molecule's trajectory (a). With geometric motion smoothing ($\tau=1$), the pink molecule moves along a rather straight trajectory (b). \rev{Trails in the first frame were added for illustration.} }
    \label{fig:smoothing}
\end{figure*}

\section{Illustrative Motion Smoothing}

Illustrative motion smoothing aims to modify the visual saliency of a cluttered dynamic visualization so that the attention is guided to the focus elements of the main story while conveying the chaotic nature of the scene. Humans are able to visually distinguish objects by their movement speed~\cite{ivry1992asymmetry}. In particular, quickly moving objects are more attention-grabbing within a set of slowly moving distractors~\cite{rosenholtz1999simple}. 
Illustrative motion smoothing utilizes this phenomenon and selectively modifies the dynamic representation of the scene so that focus elements exhibit more high-frequency motion than contextual elements. We achieve this by artificially reducing the attention-grabbing high-frequency motion in the contextual part of the dynamic scene. 
This way, the focus parts gain relatively higher visual prominence. Illustrative motion smoothing can therefore be categorized as a \emph{context suppression}~\cite{waldner2017exploring} or \emph{anti-cueing}~\cite{lowe2011cueing} technique operating on motion channels. The advantage of context suppression is that it leaves the focus elements unmodified \rev{so that the main actors of the story can be shown in full detail}. 

We exemplify illustrative motion smoothing on molecular scenes, which are especially dense and chaotic due to Brownian motion. Note that molecular animations typically show a significantly sparsified depiction of molecular environments, which are otherwise so densely packed that a dynamic visualization would be meaningless~\cite{iwasa2007visualizing}. We assume that focus elements are given from an underlying narrative, e.g., a sequence of biological structures performing reactions~\cite{kouvril2020molecumentary}. The remaining scene elements represent the context and are subject to illustrative motion smoothing. 

\subsection{Geometric Motion Smoothing}
\label{sec:geometricMotionSmoothing}

Fractal Brownian motion is a generalization of Brownian motion and is equivalent to Perlin noise~\cite{vivo2015book}. To implement Brownian motion in a real-time system, molecular positions can therefore be sampled from a continuous random noise function approximating Perlin noise. The position $p$ of the $i$'th molecule in the scene at animation time $t$ with a pre-defined fixed velocity $v$ is thereby computed as follows for each vector component $x$:  
\begin{equation}
    p_x(i, t) = n(x \cdot i + t \cdot v + n(x \cdot i + t \cdot v)), 
    \label{eq:motion}
\end{equation}
where $n$ is the continuous random noise function taking a single seed value as input. By nesting two noise functions, higher-frequency positional changes can be achieved. This results in noticeable visual jitter of the individual molecules, as shown in Figure~\ref{fig:smoothness0}. 

To achieve geometric motion smoothing, we straighten the trajectories of contextual molecules. Trajectory smoothing has been previously formulated by Le Muzic et al.~\cite{le2015illustrative} as an Infinite Impulse Response low-pass filter, which is applied to the sequence of positions obtained from the preceding simulation. For real-time implementations, where the animation sequence is computed on-the-fly, trajectories can be smoothed by reducing the influence of the nested noise function in Equation~\ref{eq:motion} by a normalized smoothing factor $\tau$: 
\begin{equation}
    p_x(i, t) = n(x \cdot i + t \cdot v + (1 - \tau) \cdot n(x \cdot i + t \cdot v)).  
    \label{eq:smoothMotion}
\end{equation}
 
By reducing the high-frequency jitter of molecular movement, geometric motion smoothing effectively shortens the trajectories (see Figure~\ref{fig:smoothness1}) and thereby \rev{reduces} the speed of context elements. In a preliminary study, Le Muzic et al.~\cite{le2015illustrative} found that by just smoothing the trajectories of focus elements, the perceived speed of the entire scene was decreased. As we apply geometric motion smoothing on the contextual scene elements, which are much more numerous than the key elements of the story in focus, we also expect to observe a considerable perceived slow-down effect. 

\subsection{Visual Motion Smoothing}
\label{sec:visualMotionSmoothing}

Visual motion smoothing adds temporal blending on contextual elements to generate a motion blur effect.
When capturing a moving object with a camera, the amount of visible motion blur depends on two aspects: the speed of the object and the camera's shutter speed~\cite{brinkmann2008art}. Within a scene, an observer can assume that the camera's shutter speed is constant. That means, to perceive more motion blur, objects need to move faster. It can therefore be expected that increasing motion blur also leads to a higher perceived speed, which has also been observed in psychophysics experiments~\cite{ivry1992asymmetry}. However, it is still unknown if the average speed of the objects in the scene is also overestimated when motion blur is only applied selectively. 

In real scenes recorded by a physical camera, visible motion blur is captured automatically as the incoming light is integrated during the camera's exposure time. In rendered scenes, motion blur needs to be artificially simulated, and multiple algorithms have been discussed~\cite{navarro2011motion}. One simple algorithm, which is suitable for real-time rendering, simulates the slower shutter speed of a camera by using an incrementally cleared accumulation buffer to create a fading motion trail behind a moving object. With this algorithm, however, more recent frames will have a higher weight resulting in a fading streak instead of a motion blur trail with averaged intensities. \rev{Such motion trails -- as also employed in the study by Höferlin et al.~\cite{hoferlin2012evaluation} -- do, however, not correspond to the behavior of real-world cameras and lead to less visual smoothing.} 

An even simpler real-time approach is to create a buffer containing the motion field of the scene between two frames. The resulting per-pixel motion vectors can then be used to apply directed Gaussian blur to the output image. While this method works well for short motion trails, it produces artifacts for longer trails due to the temporal instability of the motion vectors. As the blur can only be applied linearly along the motion vector, rapid changes in direction will result in jittery, unstable trails. To simulate real-world camera shutters more accurately, video post-processing tools provide an echo effect, which can blend the intensities of both, previous and upcoming frames. This leads to a smooth motion trail. 

Illustrative visual \rev{motion} smoothing applies motion blur to contextual elements only. In the worst case, this can introduce a disturbing visual discrepancy between artificial motion cues and the perceived movement. Ideally, if motion blur indeed causes observers to overestimate the speed, it could compensate for the lack of actual motion due to geometric motion smoothing. 

\section{Study}

We conducted a crowdsourced study \rev{with 108 users recruited from Prolific~\cite{palan2018prolific}} to answer our research questions Q1 to Q3 listed in Section~\ref{sec:introduction}. For the study, we used a molecular scene rather than more simplified stimuli to increase the engagement of the participants. 

\subsection{Hypotheses}
\label{sec:hypotheses}

With respect to comprehensiveness (\textbf{Q1}), we hypothesized that both illustrative motion smoothing techniques contribute positively to the ability to follow a story in a cluttered visualization. As targets moving with higher frequency tend to stand out among a set of slowly moving distractors~\cite{rosenholtz1999simple,ivry1992asymmetry}, \emph{geometric motion smoothing will significantly facilitate the task by making focus elements stand out from the context} (\textbf{H1.1}). In addition, blurring static context information has been shown to generate a similarly strong popout effect as traditional color highlights~\cite{kosara2002focus+}. It is, therefore, reasonable to assume that \emph{selective motion blurring of moving objects will make the non-blurred objects stand out visually} (\textbf{H1.2}).

Concerning motion perception (\textbf{Q2}), we hypothesized that, intuitively, \emph{increased geometric motion smoothing will decrease the perception of speed} (\textbf{H2.1}), as previously shown by Le Muzic et al.~\cite{le2015illustrative} when smoothing focus elements (see also Section~\ref{sec:geometricMotionSmoothing}). Conversely, previous work~\cite{vaziri2008apparent} has suggested that \emph{motion blur may generate the illusion of more quickly moving objects} (\textbf{H2.2}), as also elaborated in Section~\ref{sec:visualMotionSmoothing}. We, therefore, expected to see an interaction between motion blur and geometric motion smoothing so that \emph{increasing visual motion smoothing would compensate for the decreased speed perception due to geometric motion smoothing} (\textbf{H2.3}).  

With respect to obtrusiveness (\textbf{Q3}), Le Muzic et al.~\cite{le2015illustrative} have previously shown that smoothing the trajectories of molecules in the focus is not considered to be obtrusive but rather increased the visual appeal of the scene. We therefore also expect a similar effect when geometrically smoothing the motion of contextual element, i.e., a \emph{higher subjective visual quality of the scene with smoother context trajectories} (\textbf{H3.1}). However, long motion blur trails have been assessed as unpleasant~\cite{navarro2011perceptual} and overwhelming~\cite{hoferlin2012evaluation} by users. We, therefore, expect to observe \emph{decreased visual quality with increasing visual motion smoothing} (\textbf{H3.2}). 

\subsection{Stimuli}
\label{sec:scene}

\begin{figure}[t]
\centering
\includegraphics[width=1\columnwidth]{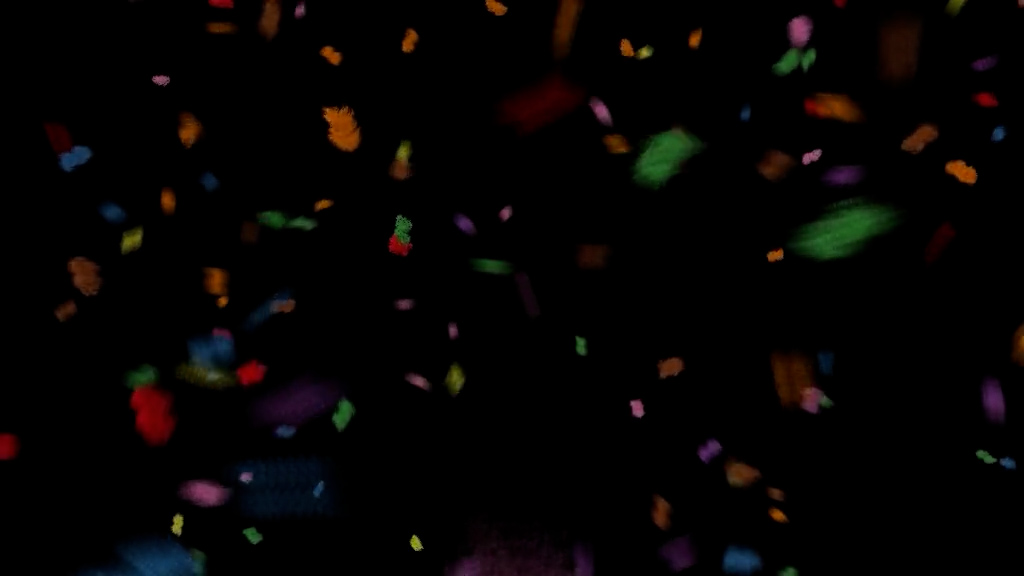}
\caption{Snapshot of the study scene with visual motion smoothing, trail length 2. The reactants in focus are here shown in red and green in the central part of the screen. }
\label{fig:stimulus}
\end{figure}

The molecular interaction scenes were presented as short pre-recorded video sequences generated with the \emph{Marion} visualization framework~\cite{mindek2017visualization}. Each video lasted 20 seconds with a resolution of 1024$\times$576 pixels. All animations comprised eight different types of molecules and 1,000 molecules in total. As clearly visible in Figure~\ref{fig:stimulus}, this does not lead to a very crowded scene. However, early informal pilot tests showed that more crowded scenes made the task to spot reactions very difficult, especially without any illustrative motion smoothing. To keep the level of frustration low, we therefore sparsified the scene. 

Molecules were color-coded according to their type \rev{so that also users without molecular biology knowledge could solve the task.} We used the 8-class Set1 color map provided by ColorBrewer~\cite{harrower2003colorbrewer}. Even though this color choice significantly lowers the visual appeal, all colors can be uniquely named (red, blue, green, purple, etc.), which facilitates remembering the reaction partners for the users. All molecules had approximately equal sizes. 

In each frame, the positions of the molecules were evaluated in real-time in the vertex shader following the Brownian motion approximation, as outlined in Section~\ref{sec:geometricMotionSmoothing}. \rev{In the geometry shader, billboard quads were used to draw one impostor sphere per atom in a molecule.} Each animation showed exactly one reaction. Reaction time was chosen at a random frame between five and ten seconds after \rev{the animation start}. A reaction sequence was composed of three stages: 1) a five seconds attraction phase where the two reacting molecules start moving towards each other, 2) a one-second reaction phase, where the molecules are physically attached and move in concordance, and 3) a five seconds repulsion phase, where the two molecules are moving away from each other again. During all three stages and eleven seconds, respectively, the two reactants represented the focus elements. No motion smoothing was applied to the focus elements during the reaction sequence. To achieve a consistent attraction motion, we blended the random movement of the two reactants with a linear interpolation moving towards a random reaction position as described by Le Muzic et al.~\cite{le2014ivm}.

Motion blur was added as a post-processing effect. To achieve smooth motion trails with adjustable lengths, a high temporal resolution is necessary. Therefore, we first rendered the animations as 120 fps videos. To add motion blur, we rendered two videos for each sequence: one containing only the reactants (i.e., the focus elements) and a second one containing the context molecules and the focus elements masked in black. Masking was necessary to resolve potential occlusions of context motion trails by the focus elements. We then simulated motion blur through an echo effect, which accumulates \rev{$n$ frames centered around the current frame into a FIFO queue. The value of $n$} depends on two parameters: 1) the ground truth speed of the scene and 2) the anticipated trail length (cf., Section~\ref{sec:design}). This allows us to control the length of the motion trails independently of the movement speed of the molecules, which was required by our study design. \rev{For each frame in the scene, the original frame was then replaced by an image consisting of the average intensities of the frames in the queue. To achieve a consistent motion trail, the beginning and the end of the animation were padded with $n/2$ frames.} In a final post-processing step, we then blended the video of the smoothed context molecules with the video containing the reactants in focus using the screen blend mode. For the study, the 120 fps videos were downsampled to 30 fps as we observed playback issues with high frame rate videos in some browsers.

\subsection{Tasks}
\label{sec:tasks}

Users were presented with ten of these short molecular animation sequences. After each animation, users had to answer the following questions: 

To test for users' ability to follow the story (\textbf{Q1}), we asked users if they could spot the reaction of two molecules in the animation. They were asked to report the colors of the two reacting molecules after the animation sequence. 
In addition, we asked users to retrospectively assess the difficulty of spotting the reaction on a continuous slider from ``impossible'' to ``very easy''. If they were not able to see the reaction, they were instructed to select two random colors and ``impossible'' \rev{on the difficulty slider}. 

To judge users' speed perception (\textbf{Q2}), we informed users in the task description that molecules move faster in warmer environments because the intensity of Brownian motion is directly proportional to the environment temperature. We then showed two video sequences without any illustrative motion smoothing representing the lowest and highest temperature that would be shown in the course of the study. These two videos served as visual calibration, and we asked users after each animation to estimate the temperature of the current scene compared to the lowest and highest temperature references shown at the beginning. This was expressed on a continuous slider from ``lowest'' to ``highest''. 

Finally, we also asked users \rev{to rate} how pleasant it was to watch the animation on a continuous slider from ``horrible'' to ``very pleasant'' to assess the obtrusiveness (\textbf{Q3}). 

In addition to the color picker, users had to adjust three continuous sliders. All three sliders were initially centered \rev{in} a neutral position, and the system would not let the users proceed unless they moved all sliders. In the color picker, there was no color selected by default, and the system would let users proceed only if they selected two colors. 

\subsection{Design}
\label{sec:design}

We used a within-subjects design with two independent variables:

\textbf{Geometric motion smoothing (GMS)} specifies if the movement of the molecules themselves is simplified. We tested two smoothness levels, namely $\tau=0$ (see Equation~\ref{eq:smoothMotion}) representing the non-smoothed scene and $\tau=1$ for the smoothed scene. Albeit a finer geometric smoothing control would be possible by choosing a $\tau$ between 0 and 1, informal early pilot tests indicated that the differences were too subtle to cause large effects on story comprehensiveness, speed perception, or obtrusiveness. 

\textbf{Visual motion smoothing (VMS)} specifies the length of the motion blur trail and thereby controls the amount of visual motion smoothing. 
We tested five different trail lengths. Trail length 0 means no motion blur is applied. Trail lengths 1 to 4 define the motion blur trail length in world space. Here, one increment in trail length roughly corresponds to the diameter of the bounding sphere of one molecule, independent of the animation speed. Note that for slow animations with high VMS trail lengths, the generated motion blur effects can be actually physically impossible as the motion trails exceed that travel period captured during the virtual shutter sequence. Figure~\ref{fig:smoothness1}, for instance, yields motion trails that would not be able to be captured by a physical camera. 

In total, each user performed 2 GMS $\times$ 5 VMS trail length $=$ 10 conditions. The order of appearance was fully randomized. We did not perform any repetitions to keep the study time for the participants as low as possible. 

For the 10 conditions, we randomly selected one out of four linearly increasing \textbf{ground truth speed} levels representing the different environment temperatures. The ground truth speed level is given as parameter $v$ as part of the seed for the random noise function (Equation~\ref{eq:smoothMotion}). The range of speed levels was empirically set to be representative of animation speeds in existing molecular animations. The maximum speed was chosen so that it was still possible to follow focus elements during their reaction. Note that molecular visualizations can never depict the true speed of molecular movement, which can be as high as 500 m/s for a molecule of gaseous oxygen at room temperature according to the Maxwell-Boltzmann distribution. Molecular animations are always just \rev{crude approximations} that try to communicate the crowding and chaos~\cite{iwasa2007visualizing}. 

As dependent variables, we obtained the following measures: to test \textbf{Q1}, we recorded the two colors users picked as reaction partners and checked if both are correct. The subjective difficulty expressed by the slider was measured on an integer scale from 0 to 100. As a measure for subjective speed (\textbf{Q2}), we used the participants' temperature estimation \rev{relative to the reference videos}, which was \rev{stored} on a scale from 0 to 100. Finally, the obtrusiveness (\textbf{Q3}) was expressed through users' assessment of pleasure watching the animation on a scale from 0 to 100, where 50 can be considered a neutral response.

\subsection{Procedure and Apparatus}
\label{sec:procedure}

The study was hosted on a university server. Users were recruited from Prolific~\cite{palan2018prolific} and forwarded to the university server URL. We used a custom in-house web service, which takes care of randomizing the stimuli and logging user responses. 

We first showed users a welcome page with a task description, including the two calibration videos for the minimum and maximum temperature, respectively (see Section~\ref{sec:tasks}). For both videos, we also asked users to try and spot the reactions. To help them understand the task, we mentioned the colors of the reaction partners, the video time when the reaction starts, and the approximate trajectory they take during the reaction. After assessing these two calibration videos, we let users perform a test run. This test run video had VMS with trail length 4 and no GMS. As reaction partners, we chose red and green as an implicit color blindness test. Users received feedback on their responses to the test task and could repeatedly adjust their input. 
We kindly asked users to return their submission if they could not find the matching colors of the reaction partners for this test run. 

On the recruiting platform, we informed users that the study would ask for a color assessment and that the videos would include many rapidly moving elements. We also informed users that we would check if their browser window is at least 1024 pixels wide. Furthermore, we used cookies to test if users navigate back or refresh the page and informed users that doing so would invalidate their assignment. Finally, we asked users to kindly return their submission if the videos were not playing smoothly or would show artifacts. All users who finished all ten stimulus responses were presented with a valid completion code and were rewarded for their participation. 

After the welcome and instruction page, users filled out a short demographic questionnaire asking their age, sex, and self-assessed knowledge about molecular biology. Before each stimulus display, users could take a rest and proceed on button press. After pressing the button, the 20-second animation started playing without any video controls. After finishing the video, the questionnaire, as described in Section~\ref{sec:tasks}, was displayed.

\subsection{Recruiting}

An a-priori power analysis yielded a sample size of 80 to show large effects (i.e., Cohen's $f \geq 0.4$) with power $1 - \beta = .8$ and $\alpha=.05$. We, therefore, chose 120 participants as our target sample size to reliably show large effects. Within the recruiting platform, we opted to balance female and male participants. \rev{Since edutainment animations are usually targeted towards a broad audience, we did not filter participants further based on Prolific's available filters, such as age or nationality, to have a diverse group. }
Our pilot study with nine participants recruited in a local environment showed that users required 10 to maximum 15 minutes to complete the task. We, therefore, paid all participants who completed the assignment 2.25 \pounds, which is Prolific's recommended payment for a 15-minute assignment.  

\section{Results}

The nine users of our pilot test could correctly spot, on average, 70\% of the reactions. The correlation between their estimated speed and the ground truth speed was high with an average $r=0.74$. As inclusion criteria, we therefore determined that participants should report at least three correct reaction pairs and achieve at least a medium correlation between their speed estimation and the ground truth speed. 

In total, 123 users participated in the study \rev{of} which four did not finish all trials and were therefore excluded by the recruiting platform. From the remaining 119 users, 11 did not fulfill our inclusion criteria and were rewarded, but excluded from further analyses. 
Of the remaining 108 participants, 47\% were female. Ages ranged from 18 to 60 with a mean age of 26.9; one person chose not to state their age. The mean reported experience level in the field of molecular biology was expectantly low with 1.83 out of a range between 1 and 5. Eight participants stated that they are experienced, \rev{and} no participant claimed to be an expert. 

Before running the statistical analyses, we performed normality tests on the obtained measures. Since most of the normality tests failed, we employed non-parametric tests. For those, responses were averaged per level of the main effect to be investigated per user. To visualize the results, we use letter-value plots~\cite{letter-value-plot}. All pairwise post-hoc comparisons were Bonferroni-adjusted.

\subsection{Q1: Comprehensiveness}

\begin{figure}[ht]
\centering
\includegraphics[width=0.9\columnwidth]{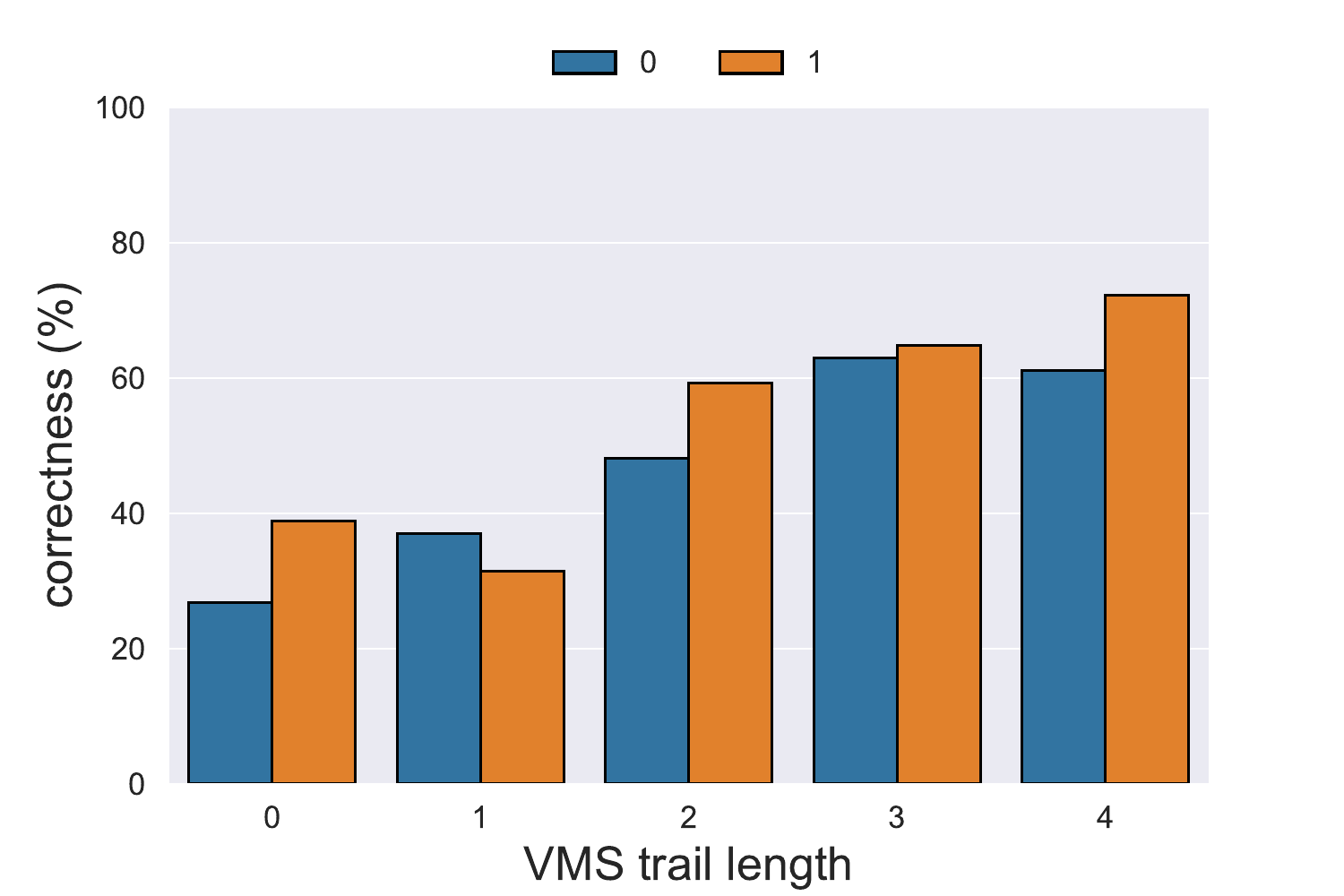}
\caption{Overall percentage of correctly identified reaction pairs dependent on VMS trail length, grouped by GMS (non-smoothed $\tau=0$ in blue and smooth $\tau=1$ in orange). }
\label{fig:tl_t_re_bin}
\end{figure}

In our study, users could correctly identify 27\% of the reactions without any illustrative motion smoothing (see leftmost blue bar in Figure~\ref{fig:tl_t_re_bin}). This, therefore, represents the lower bound for comprehensiveness in our study. 
\rev{Ground truth speed had a strong influence on the percentage of correct responses ($R^2=.95$), while increasing luminance of one of the reaction partners did not ($R^2=.04$). }

\begin{figure}[ht]
\centering
\includegraphics[width=0.9\columnwidth]{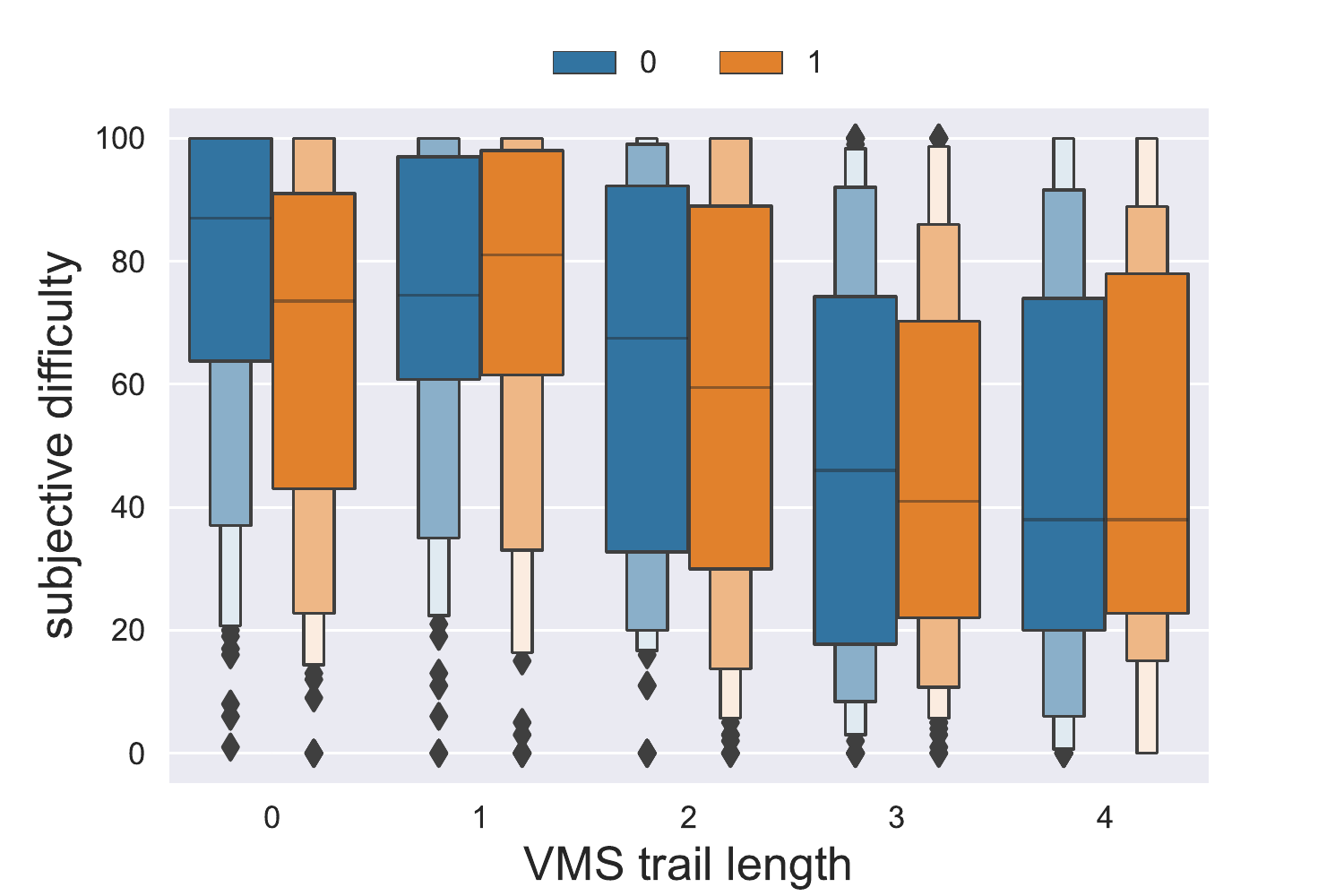}
\caption{How difficult it was considered by users to spot a reaction dependent on VMS trail length, grouped by GMS. }
\label{fig:tl_t_di}
\end{figure}

Across all levels of VMS trail length, 47.2\% of the responses were correct without GMS; with GMS, it was 53.3\% (compare blue and orange bars in Figure~\ref{fig:tl_t_re_bin}). A \rev{Wilcoxon Signed Rank} test shows that this difference is not statistically significant: \rev{$Z=-1.890;p=.059$}.  
However, users consider the task significantly easier with GMS: $Z=2.477;p=.013$. Without GMS, the median difficulty of the task was rated with 87; with smoothing, the difficulty decreased slightly to 73.5 (see blue and orange boxes in Figure~\ref{fig:tl_t_di}). \textbf{H1.1} is therefore only partially supported: \emph{GMS makes it subjectively easier to spot reactions, but this subjective impression is not backed up by objective task performance measures}.  

For stimuli without VMS, only 32.9\% of the responses were correct. The percentage of correctly reported reaction pairs increased with every trajectory length step up to 66.67\% (see groups of bars in Figure~\ref{fig:tl_t_re_bin}). A Friedman test confirmed that this increase is significant: 
 $\chi^2(4)=81.724;p<.001$. According to Bonferroni-corrected post-hoc comparisons, moderate VMS with a trail length of 2, as shown in Figure~\ref{fig:stimulus}, already leads to significantly fewer errors compared to having no VMS. 
 
The subjective difficulty of the task decreased from a median value of 87 with trail length 0 to 38 with trail length 4, as illustrated in Figure~\ref{fig:tl_t_di}. According to a Friedman test, this decrease is statistically significant: $\chi^2(4)=92.204;p<.001$. Similarly to the objectively measured reaction errors, a trail length of 2 already yields significantly easier task execution. There is no more statistically significant difference in terms of subjective ease of task execution when further extending the trails. Thus, \textbf{H1.2} is supported: \emph{moderate visual motion smoothing can facilitate the task to spot reactions -- both subjectively and objectively. However, extending the trail length further does not necessarily make the task easier.}

\subsection{Q2: Speed Perception}

Across all smoothed and unsmoothed conditions, users gave significantly different speed estimations based on the ground truth speed. This shows that users took the task seriously. 
The baseline speed estimation without any motion smoothing effect in our study can be approximated using the following linear regression: 
\begin{equation}
    es = 30.4 + 0.6s, 
    \label{eq:tl0_n_es}
\end{equation}
where $es$ is the estimated speed and $s$ is the ground truth speed in percentage from minimum (i.e., ground truth speed level 1) to maximum speed (ground truth speed level 4).

\begin{figure}[ht]
\centering
\includegraphics[width=0.9\columnwidth]{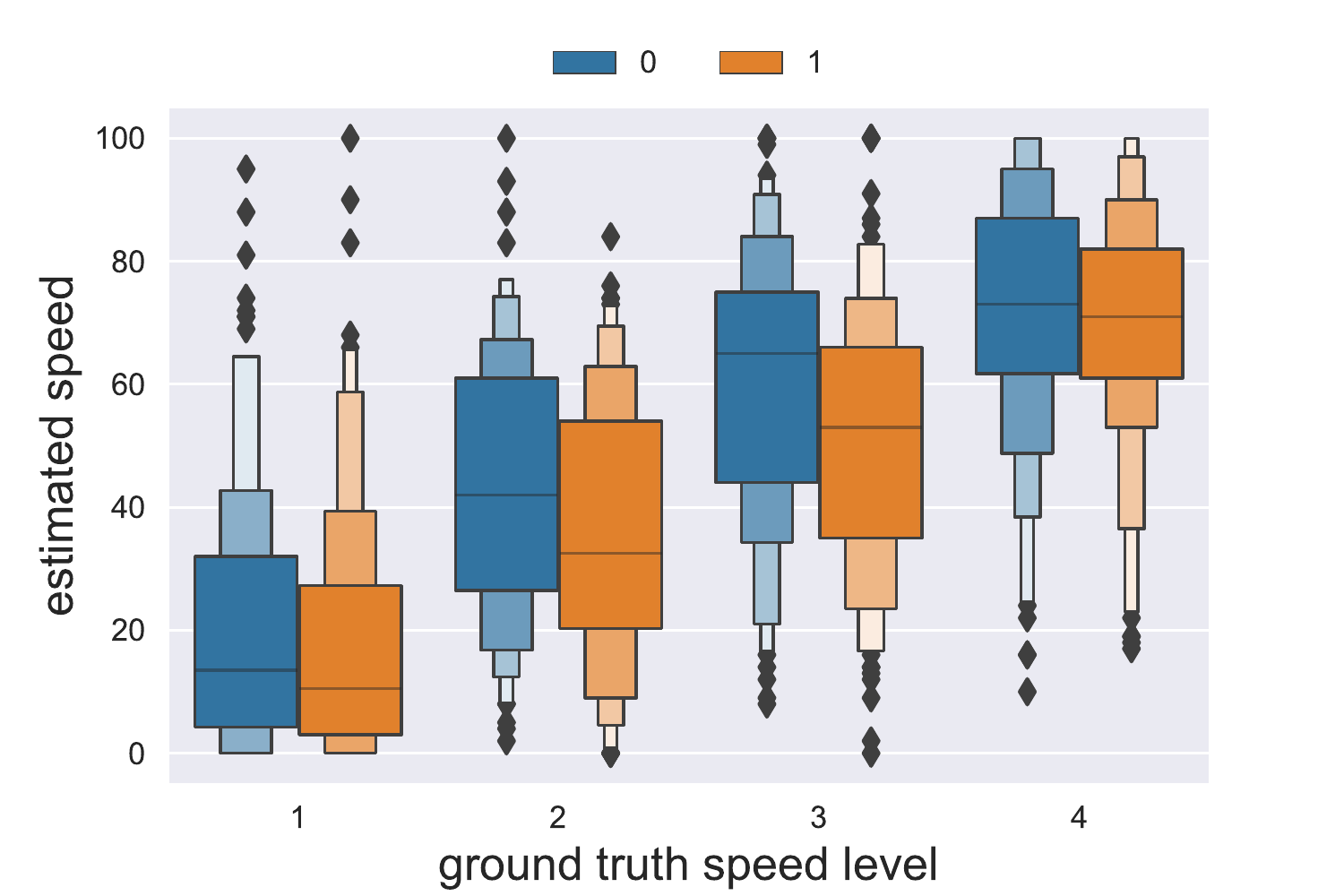}
\caption{Estimated speed dependent on ground truth speed, grouped by GMS. }
\label{fig:s_t_es}
\end{figure}

A Wilcoxon signed rank test showed that adding GMS has a significant effect on the users' estimated speed: $Z=-3.094; p=.002$. As illustrated by the orange bars in Figure~\ref{fig:s_t_es}, users' speed estimation is consistently lower in the presence of GMS. 
This means that \textbf{H2.1} is supported: \emph{GMS significantly slows down the perceived speed of the animation}.

Considering only stimuli responses with smoothed trajectories and with VMS trail length 0, we can model speed estimation under GMS using the following linear regression: 
\begin{equation}
    es = 19.3 + 0.6s. 
\end{equation}
We can see that the regression coefficient is identical to the baseline regression (Equation~\ref{eq:tl0_n_es}). The lower intercept confirms that GMS is leading to a constantly lower speed estimation compared to the non-smoothed baseline, irrespective of the ground truth speed. This means that the effect of GMS, as it was implemented in our study, could be compensated by adding a small constant speed factor.  

\begin{figure}[ht]
\centering
\includegraphics[width=0.9\columnwidth]{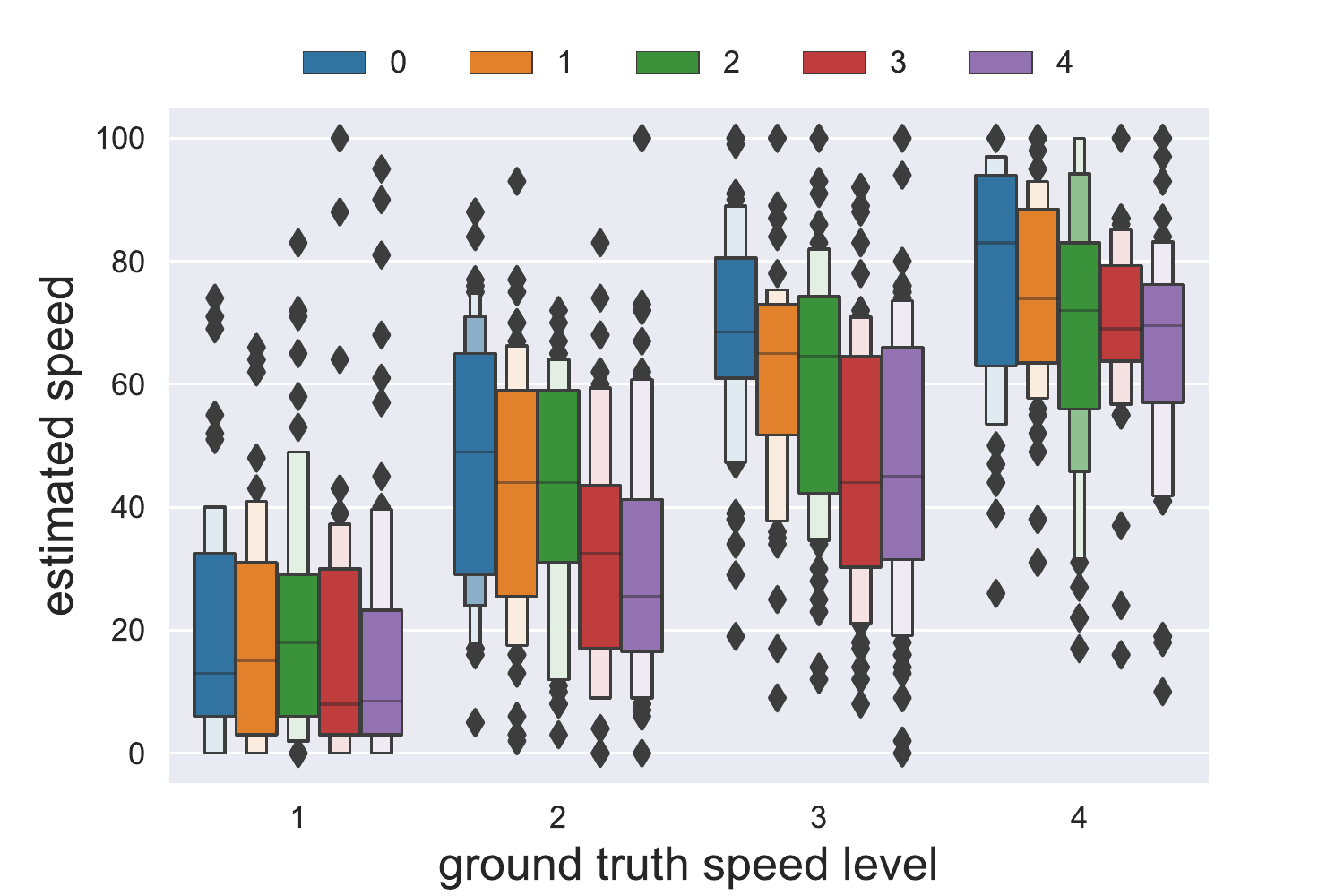}
\caption{Estimated speed dependent on ground truth speed, grouped by increasing VMS trail length. }
\label{fig:s_tl_es}
\end{figure}

Adding VMS to contextual scene elements also had a significant impact on the perceived speed according to a Friedman test: $\chi^2(4)=27.914;p<.001$.
Figure~\ref{fig:s_tl_es} shows that, contrary to our expectations, the perceived speed \emph{decreases} with increased VMS trail length -- at least for faster animations. 
Post-hoc comparisons revealed that speed underestimation reaches significance with trail length 2, but does not change significantly with longer trails. Thus, \textbf{H2.2} is not supported: \emph{on the contrary, adding VMS with moderate trail lengths leads to a significantly slower speed perception}. 

Since moderate VMS with a trail length of 2 seems to be a sweet spot concerning story comprehensiveness and speed perception, we look at all responses gathered for stimuli with a trail length of 2 and non-smoothed trajectories. Users' speed estimations for these stimuli can be approximated by the following linear regression: 
\begin{equation}
    es = 29.8 + 0.4s. 
    \label{eq:tl2_n_es}
\end{equation}
As the intercept is similar to the baseline regression in Equation~\ref{eq:tl0_n_es}, this confirms that speed perception is comparable to the baseline for lower speeds despite VMS. However, as the ground truth speed increases, the perceived speed does not seem to increase as strongly in the presence of VMS. By substituting $es$ in Equation~\ref{eq:tl0_n_es} with the regression term in Equation~\ref{eq:tl2_n_es} and solving for the ground truth speed of the blurred scene, we obtain a linear formula to compute a compensated animation speed: $cs = 1.5(s + 1)$. That means the moderate motion blur in our study could be compensated by increasing the ground truth speed by a factor of approximately 50\%. 

As the normality assumption is violated and we need to perform non-parametric tests, we are not able to statistically test for interaction effects. However, since both illustrative motion smoothing techniques effectively lead to an underestimation of the animation speed, our expected interaction effect is not possible to observe anyway. \rev{Therefore, \textbf{H2.3} is not supported:} \emph{both, geometric and visual motion smoothing, lead to slower speed perception and thus cannot compensate for each other.}

\subsection{Q3: Obtrusiveness}

\begin{figure}[ht]
\centering
\includegraphics[width=0.9\columnwidth]{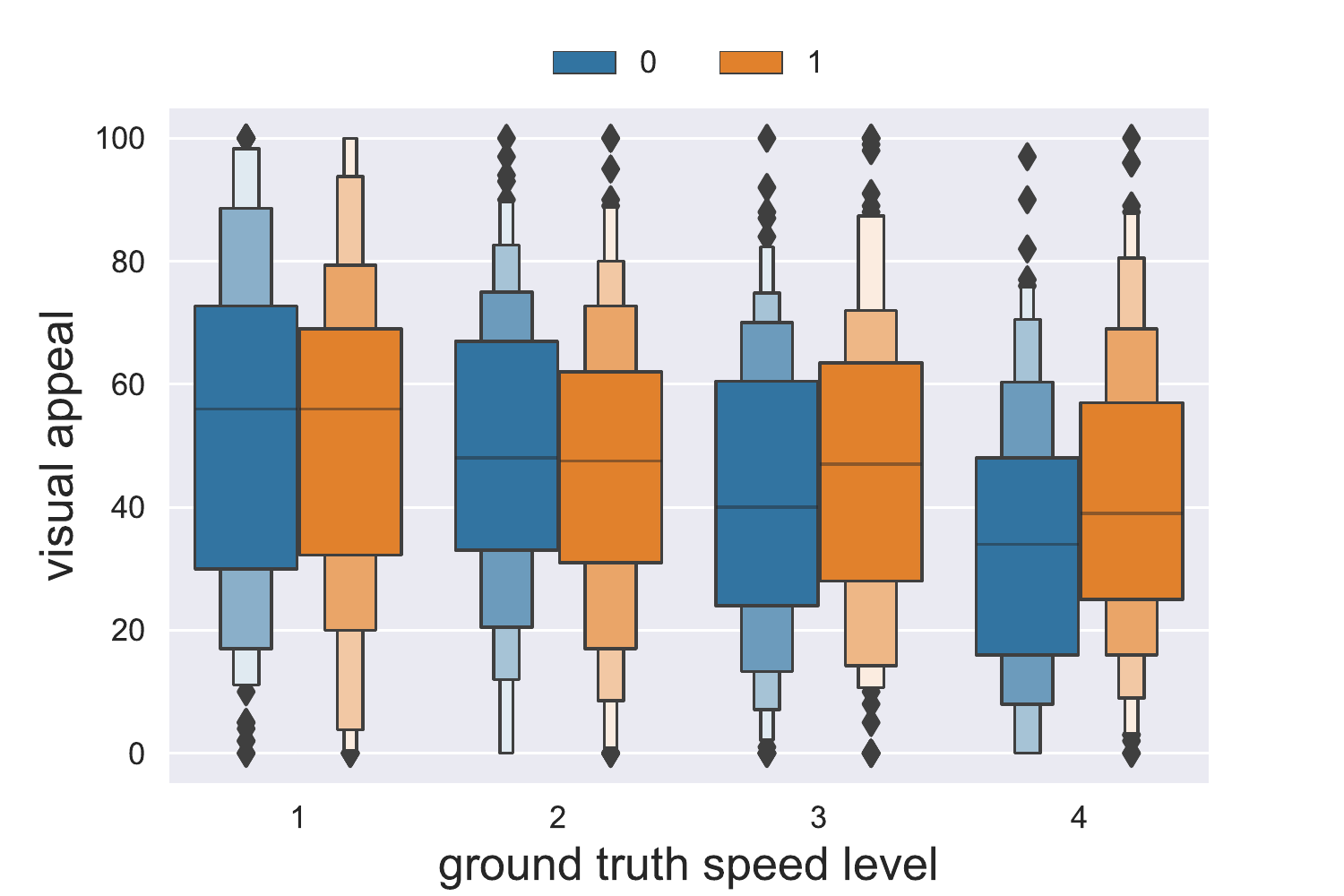}
\caption{Subjective visual appeal dependent on ground truth speed, grouped by GMS (50 is neutral). }
\label{fig:gtsl_GMS_ae}
\end{figure}

We can observe a slight increase in median aesthetics judgments when adding GMS (from 42.5 to 47), as indicated by the slightly higher median values in the orange boxes of Figure~\ref{fig:gtsl_GMS_ae}. This difference is statistically significant according to a Wilcoxon Signed-Rank test: $Z=2.496;p=.013$. \textbf{H3.1} is thereby supported: \emph{geometric motion smoothing has a small positive effect on perceived scene aesthetics}. 

\begin{figure}[ht]
\centering
\includegraphics[width=0.9\columnwidth]{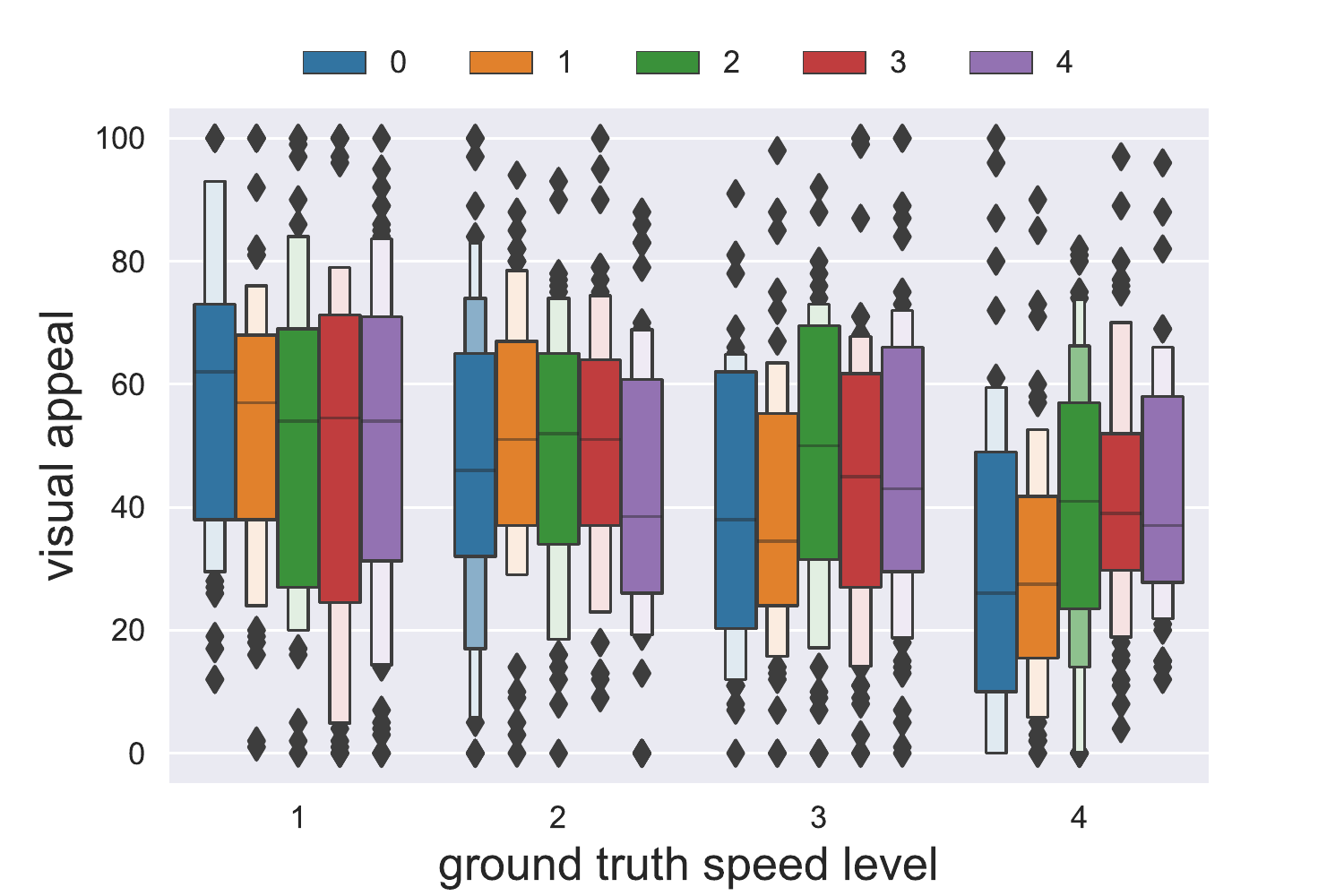}
\caption{Subjective visual appeal dependent on ground truth speed, grouped by VMS trail length (50 is neutral). }
\label{fig:gtsl_VMS_ae}
\end{figure}

Aesthetic judgments were not following VMS trail length: the highest median aesthetics judgment was found for trail length 2 (47.5). The lowest values were issued for scenes without or only very decent motion blur (i.e., trail lengths 0 and 1) with a median rating of 44, and the longest trails with length 4 with 43.5. However, as can be seen in Figure~\ref{fig:gtsl_VMS_ae}, these differences are small -- at least for low ground truth speed levels. According to a Friedman test visual appeal ratings did not change significantly with VMS trail length: $\chi^2=7.184;p=.126$. This means \textbf{H3.2} \rev{is not supported}: \emph{visual motion smoothing is neither considered obtrusive nor does it increase the visual appeal of the scene. }

\section{Discussion}

Our study showed that illustrative motion smoothing can guide the attention towards selected actors in a crowded and chaotic scene. Adding illustrative motion smoothing effects thereby has little influence on the visual appeal of the scene. However, the attention guidance effect comes at the cost of slower scene perception. 

Users generally seem to like \textbf{geometric motion smoothing}. They have the impression that the task becomes \rev{slightly} easier when smoothing of trajectories is applied, and they think that the scene looks a bit more appealing. Like visual motion smoothing, geometric motion smoothing also perceptually slows down the scene. \rev{Overall, we can observe a moderate correlation between estimated speed and subjective task difficulty ($r=.314$) and a weak negative correlation between estimated speed and visual appeal ($r-.226$). The perceived slow-down effect due to geometric motion smoothing could therefore at least partially explain the lower subjective task load and higher aesthetics ratings.}

With our implementation of geometric motion smoothing, the high-frequency jitter of the unsmoothed focus elements was not intensive enough to generate a sufficiently strong motion popout effect~\cite{rosenholtz1999simple} from the smoothly moving context elements. 
Overall, geometric motion smoothing of contextual elements alone is probably not sufficient to guide the users' attention in highly cluttered environments. 

Moderate \textbf{visual motion smoothing} significantly improves the users' ability to follow the story -- objectively and subjectively. At the same time, it also significantly slows down the animation for the users. This was not necessarily expected from prior work, which showed mixed results. To the best of our knowledge, Holm et al.~\cite{holm2016increasing} presented the only study that found an unexpected underestimation of camera velocity when motion blur was applied to the surrounding scene. One plausible explanation is the fact that motion blur decreases the contrast of context elements considerably (see Figure~\ref{fig:teaser}): It has been shown that lower-contrast objects appear to move slower than high-contrast ones~\cite{stone1992human}. 

Generally, we can conclude that having extensively long motion trails does not increase the ability to follow the story or the negative effect on speed perception; the effects seem to be rather saturated with a moderate trail length. 
Especially the subjective visual appeal seems to slightly decrease with increasing trail length. Similarly, Navarro et al.~\cite{navarro2011perceptual} have found that motion blur can be perceived as too excessive at some point. Indeed, if motion trails are getting too long, they are no longer physically plausible, which may cause irritation for the observers. 

\section{Design Guidelines}

From the findings of our study, we derived design guidelines for animators of complex dynamic visualizations: 

\textbf{Do not overdo visual motion smoothing. } 
Similarly, as a previous study on a simple scene has indicated~\cite{navarro2011perceptual}, adding extensive motion trails, like shown in Figure~\ref{fig:teaser_tl4}, may be perceived as too excessive and does not add any measurable benefit for story comprehensiveness. Visual motion smoothing by adding short motion trails as shown in Figure~\ref{fig:teaser_tl2} can already facilitate story comprehension.  

\textbf{Increase the animation speed to compensate for unwanted slow-down effects. }
Illustrative motion smoothing leads to an underestimation of the animation's speed. If animators seek to communicate the chaotic nature of the scene while using illustrative motion smoothing to guide the user's attention, they should increase the speed of the animation to compensate for this slow-down effect. However, keeping the pace up constantly may not be desirable: 

\textbf{Avoid constantly fast animations. } 
We could observe a decrease in story comprehension and subjective visual appeal with increasing animation speed. It may therefore be advisable to relax the requirement to communicate the chaotic nature of a scene at all times. Reducing the animation speed can facilitate learning of complex interactions~\cite{tversky2002animation,rekik2020decreasing}, while also enhancing the overall viewing experience with added enjoyment. 

\textbf{Smooth the trajectories of contextual elements to perceptually lower the pace of very fast animations.} 
Geometric motion smoothing is visually appealing and subjectively improves the ability to follow a story. It could therefore be a useful method to subjectively decrease the animation speed while still being able to show the complex movement trajectories of focus elements in full detail. 

\textbf{Add further visual cues to guide the attention reliably. }
Illustrative motion smoothing alone could not fully reliably attract users' attention to the main actors of the story. The success rate rather peaked at 60-70\% (see Figure~\ref{fig:tl_t_re_bin}). Further attention cues are necessary to further increase the saliency of focus elements, such as lowlighting of contextual scene elements~\cite{de2007attention}, halos around focus elements, or camera movement following the main actors of the story~\cite{iwasa2007visualizing}.  

\section{Limitations and Future Work}

In this work, we studied two methods to generate illustrative motion smoothing in a highly specialized application domain. Geometric motion smoothing is an effect that is only reasonably applicable for dynamic visualizations of biologically inspired motion, such as Brownian motion or crowd simulations. However, visual motion smoothing is more generally applicable and could be studied in different contexts in the future. 

Even within the same application domain, there are several factors that have been controlled for in this study, which may also have an impact on story comprehensiveness, speed perception, and visual appeal. Examples are \rev{the crowding} of the scene and the heterogeneity of the molecules in terms of size and shape, which will further increase the complexity of the animation. It can be assumed that illustrative motion smoothing will also help to decrease the visual complexity of denser and more chaotic scenes.

Visual motion smoothing was found to be effective in terms of story comprehensiveness, but it was generated in a post-processing step for this study. Real-time motion blur algorithms exist but may lead to lower visual quality effects. Real-time motion blur may therefore yield lower visual appeal or influence speed perception in a different way. 

Finally, illustrative motion smoothing was studied as a single method to guide the user's attention in this paper. Manipulating multiple visual channels to generate an attention guidance effect can be highly effective~\rev{\cite{waldner2017exploring,berry2009malaria,berry2018dna}} and may help to overcome the limited visual prominence of motion smoothing effects alone. Investigating the interaction of multiple visual cues\rev{, such as color or static depth-of-field effects~\cite{kosara2002focus+},} in such highly complex dynamic visualizations is therefore also important future work. 

\rev{\section*{Acknowledgments}

The authors wish to acknowledge all anonymous participants of this study. The authors acknowledge TU Wien Bibliothek for financial support through its Open Access Funding Programme. 
}
\bibliographystyle{eg-alpha-doi}  
\bibliography{bib.bib}        

\newcommand{\etalchar}[1]{$^{#1}$}
\begin{thebibliography}{\uppercase{LMWPV15}}

\bibitem[ADA{\etalchar{*}}04]{agarwala2004interactive}
\textsc{Agarwala A., Dontcheva M., Agrawala M., Drucker S., Colburn A., Curless
  B., Salesin D., Cohen M.}:
\newblock Interactive digital photomontage.
\newblock In \emph{ACM SIGGRAPH 2004 Papers}. 2004, pp.~294--302.

\bibitem[Ber09]{berry2009malaria}
\textsc{Berry D.}:
\newblock The malaria lifecycle.
\newblock SIGGRAPH '09, Association for Computing Machinery, p.~91.

\bibitem[Ber18]{berry2018dna}
\textsc{Berry D.}:
\newblock {DNA animations for science-art exhibition}, 2018.
\newblock URL: \url{https://www.youtube.com/watch?v=7Hk9jct2ozY}.

\bibitem[Bri08]{brinkmann2008art}
\textsc{Brinkmann R.}:
\newblock \emph{The art and science of digital compositing: Techniques for
  visual effects, animation and motion graphics}.
\newblock Morgan Kaufmann, 2008.

\bibitem[BSM{\etalchar{*}}13]{booth2013guiding}
\textsc{Booth T., Sridharan S., McNamara A., Grimm C., Bailey R.}:
\newblock Guiding attention in controlled real-world environments.
\newblock In \emph{Proceedings of the ACM Symposium on Applied Perception}
  (2013), pp.~75--82.

\bibitem[BTM{\etalchar{*}}19]{byvska2019analysis}
\textsc{By{\v{s}}ka J., Trautner T., Marques S.~M., Damborsk{\`y} J.,
  Kozl{\'\i}kov{\'a} B., Waldner M.}:
\newblock Analysis of long molecular dynamics simulations using interactive
  focus+ context visualization.
\newblock In \emph{Computer Graphics Forum} (2019), vol.~38, Wiley Online
  Library, pp.~441--453.

\bibitem[BTV14]{birkeland2014perceptually}
\textsc{Birkeland {\AA}., Turkay C., Viola I.}:
\newblock Perceptually uniform motion space.
\newblock \emph{IEEE Transactions on Visualization and Computer Graphics 20},
  11 (2014), 1542--1554.

\bibitem[Bur80]{burr1980motion}
\textsc{Burr D.}:
\newblock Motion smear.
\newblock \emph{Nature 284}, 5752 (1980), 164--165.

\bibitem[BZOP07]{bouvier2007motion}
\textsc{Bouvier-Zappa S., Ostromoukhov V., Poulin P.}:
\newblock Motion cues for illustration of skeletal motion capture data.
\newblock In \emph{Proceedings of the 5th international symposium on
  Non-photorealistic animation and rendering} (2007), pp.~133--140.

\bibitem[CNM96]{chen1996image}
\textsc{Chen W.-G., Nandhakumar N., Martin W.~N.}:
\newblock Image motion estimation from motion smear-a new computational model.
\newblock \emph{IEEE Transactions on Pattern Analysis and Machine Intelligence
  18}, 4 (1996), 412--425.

\bibitem[CRP{\etalchar{*}}16]{chevalier2016animations}
\textsc{Chevalier F., Riche N.~H., Plaisant C., Chalbi A., Hurter C.}:
\newblock Animations 25 years later: New roles and opportunities.
\newblock In \emph{Proceedings of the International Working Conference on
  Advanced Visual Interfaces} (2016), pp.~280--287.

\bibitem[DKTRP07]{de2007attention}
\textsc{De~Koning B.~B., Tabbers H.~K., Rikers R.~M., Paas F.}:
\newblock Attention cueing as a means to enhance learning from an animation.
\newblock \emph{Applied Cognitive Psychology: The Official Journal of the
  Society for Applied Research in Memory and Cognition 21}, 6 (2007), 731--746.

\bibitem[FLS08]{fischer2008effects}
\textsc{Fischer S., Lowe R.~K., Schwan S.}:
\newblock Effects of presentation speed of a dynamic visualization on the
  understanding of a mechanical system.
\newblock \emph{Applied Cognitive Psychology: The Official Journal of the
  Society for Applied Research in Memory and Cognition 22}, 8 (2008),
  1126--1141.

\bibitem[GMF{\etalchar{*}}21]{garrison2021exploration}
\textsc{Garrison L., Meuschke M., Fairman J.~E., Smit N.~N., Preim B., Bruckner
  S.}:
\newblock An exploration of practice and preferences for the visual
  communication of biomedical processes.
\newblock In \emph{VCBM} (2021), pp.~1--12.

\bibitem[HB03]{harrower2003colorbrewer}
\textsc{Harrower M., Brewer C.~A.}:
\newblock Colorbrewer. org: an online tool for selecting colour schemes for
  maps.
\newblock \emph{The Cartographic Journal 40}, 1 (2003), 27--37.

\bibitem[HKH{\etalchar{*}}12]{hoferlin2012evaluation}
\textsc{H{\"o}ferlin M., Kurzhals K., H{\"o}ferlin B., Heidemann G., Weiskopf
  D.}:
\newblock Evaluation of fast-forward video visualization.
\newblock \emph{IEEE Transactions on Visualization and Computer Graphics 18},
  12 (2012), 2095--2103.

\bibitem[HKW11]{letter-value-plot}
\textsc{Hofmann H., Kafadar K., Wickham H.}:
\newblock \emph{Letter-value plots: Boxplots for large data}.
\newblock Tech. rep., had.co.nz, 2011.

\bibitem[HSK16]{holm2016increasing}
\textsc{Holm K.~L., Skovhus N., Kraus M.}:
\newblock Increasing the perceived camera velocity in 3d racing games by
  changing camera attributes.
\newblock In \emph{Interactivity, Game Creation, Design, Learning, and
  Innovation}. Springer, 2016, pp.~121--128.

\bibitem[IC92]{ivry1992asymmetry}
\textsc{Ivry R.~B., Cohen A.}:
\newblock Asymmetry in visual search for targets defined by differences in
  movement speed.
\newblock \emph{Journal of Experimental Psychology: Human Perception and
  Performance 18}, 4 (1992), 1045.

\bibitem[IDP03]{itti2003realistic}
\textsc{Itti L., Dhavale N., Pighin F.}:
\newblock Realistic avatar eye and head animation using a neurobiological model
  of visual attention.
\newblock In \emph{Applications and Science of Neural Networks, Fuzzy Systems,
  and Evolutionary Computation VI} (2003), vol.~5200, International Society for
  Optics and Photonics, pp.~64--78.

\bibitem[Itt05]{itti2005quantifying}
\textsc{Itti L.}:
\newblock Quantifying the contribution of low-level saliency to human eye
  movements in dynamic scenes.
\newblock \emph{Visual Cognition 12}, 6 (2005), 1093--1123.

\bibitem[Iwa07]{iwasa2007visualizing}
\textsc{Iwasa J.}:
\newblock Visualizing the origins of life: molecular animation for scientific
  research and education.
\newblock In \emph{ACM SIGGRAPH 2007 educators program}. 2007, pp.~21--es.

\bibitem[JAG{\etalchar{*}}11]{johnson2011epmv}
\textsc{Johnson G.~T., Autin L., Goodsell D.~S., Sanner M.~F., Olson A.~J.}:
\newblock epmv embeds molecular modeling into professional animation software
  environments.
\newblock \emph{Structure 19}, 3 (2011), 293--303.

\bibitem[Jen17]{jenkinson2017role}
\textsc{Jenkinson J.}:
\newblock The role of craft-based knowledge in the design of dynamic
  visualizations.
\newblock In \emph{Learning from Dynamic Visualization}. Springer, 2017,
  pp.~93--117.

\bibitem[JM12]{jenkinson2012visualizing}
\textsc{Jenkinson J., McGill G.}:
\newblock Visualizing protein interactions and dynamics: evolving a visual
  language for molecular animation.
\newblock \emph{CBE—Life Sciences Education 11}, 1 (2012), 103--110.

\bibitem[JR05]{joshi2005illustration}
\textsc{Joshi A., Rheingans P.}:
\newblock Illustration-inspired techniques for visualizing time-varying data.
\newblock In \emph{VIS 05. IEEE Visualization, 2005.} (2005), IEEE,
  pp.~679--686.

\bibitem[KKF{\etalchar{*}}17]{kozlikova2017visualization}
\textsc{Kozl{\'\i}kov{\'a} B., Krone M., Falk M., Lindow N., Baaden M., Baum
  D., Viola I., Parulek J., Hege H.-C.}:
\newblock Visualization of biomolecular structures: State of the art revisited.
\newblock In \emph{Computer Graphics Forum} (2017), vol.~36, Wiley Online
  Library, pp.~178--204.

\bibitem[KMH02]{kosara2002focus+}
\textsc{Kosara R., Miksch S., Hauser H.}:
\newblock Focus+ context taken literally.
\newblock \emph{IEEE Computer Graphics and Applications 22}, 1 (2002), 22--29.

\bibitem[KSM{\etalchar{*}}23]{kouvril2020molecumentary}
\textsc{Kou{\v{r}}il D., Strnad O., Mindek P., Halladjian S., Isenberg T.,
  Gr{\"o}ller M.~E., Viola I.}:
\newblock Molecumentary: Adaptable narrated documentaries using molecular
  visualization.
\newblock \emph{IEEE Transactions on Visualization and Computer Graphics 29}, 3
  (2023), 1733--1747.

\bibitem[LB11]{lowe2011cueing}
\textsc{Lowe R., Boucheix J.-M.}:
\newblock Cueing complex animations: Does direction of attention foster
  learning processes?
\newblock \emph{Learning and Instruction 21}, 5 (2011), 650--663.

\bibitem[LCL19]{lee2019simulating}
\textsc{Lee E.-C., Cho Y.-H., Lee I.-K.}:
\newblock Simulating water resistance in a virtual underwater experience using
  a visual motion delay effect.
\newblock In \emph{2019 IEEE Conference on Virtual Reality and 3D User
  Interfaces (VR)} (2019), IEEE, pp.~259--266.

\bibitem[LGP14]{lawonn2014adaptive}
\textsc{Lawonn K., Gasteiger R., Preim B.}:
\newblock Adaptive surface visualization of vessels with animated blood flow.
\newblock In \emph{Computer Graphics Forum} (2014), vol.~33, Wiley Online
  Library, pp.~16--27.

\bibitem[LKEP14]{lawonn2014line}
\textsc{Lawonn K., Krone M., Ertl T., Preim B.}:
\newblock Line integral convolution for real-time illustration of molecular
  surface shape and salient regions.
\newblock In \emph{Computer Graphics Forum} (2014), vol.~33, Wiley Online
  Library, pp.~181--190.

\bibitem[LMPSV14]{le2014ivm}
\textsc{Le~Muzic M., Parulek J., Stavrum A., Viola I.}:
\newblock Illustrative visualization of molecular reactions using omniscient
  intelligence and passive agents.
\newblock \emph{Computer Graphics Forum 33}, 3 (2014), 141--150.

\bibitem[LMWPV15]{le2015illustrative}
\textsc{Le~Muzic M., Waldner M., Parulek J., Viola I.}:
\newblock Illustrative timelapse: A technique for illustrative visualization of
  particle-based simulations.
\newblock In \emph{2015 IEEE Pacific Visualization Symposium (PacificVis)}
  (2015), IEEE, pp.~247--254.

\bibitem[LRB{\etalchar{*}}16]{langbehn2016visual}
\textsc{Langbehn E., Raupp T., Bruder G., Steinicke F., Bolte B., Lappe M.}:
\newblock Visual blur in immersive virtual environments: Does depth of field or
  motion blur affect distance and speed estimation?
\newblock In \emph{Proceedings of the 22nd ACM Conference on Virtual Reality
  Software and Technology} (2016), pp.~241--250.

\bibitem[MI13]{meghdadi2013interactive}
\textsc{Meghdadi A.~H., Irani P.}:
\newblock Interactive exploration of surveillance video through action shot
  summarization and trajectory visualization.
\newblock \emph{IEEE Transactions on Visualization and Computer Graphics 19},
  12 (2013), 2119--2128.

\bibitem[MKS{\etalchar{*}}17]{mindek2017visualization}
\textsc{Mindek P., Kou{\v{r}}il D., Sorger J., Toloudis D., Lyons B., Johnson
  G., Gr{\"o}ller M.~E., Viola I.}:
\newblock Visualization multi-pipeline for communicating biology.
\newblock \emph{IEEE Transactions on Visualization and Computer Graphics 24}, 1
  (2017), 883--892.

\bibitem[MSHH11]{mital2011clustering}
\textsc{Mital P.~K., Smith T.~J., Hill R.~L., Henderson J.~M.}:
\newblock Clustering of gaze during dynamic scene viewing is predicted by
  motion.
\newblock \emph{Cognitive computation 3}, 1 (2011), 5--24.

\bibitem[MYY{\etalchar{*}}10]{mitra2010illustrating}
\textsc{Mitra N.~J., Yang Y.-L., Yan D.-M., Li W., Agrawala M.}:
\newblock Illustrating how mechanical assemblies work.
\newblock \emph{ACM Transactions on Graphics-TOG 29}, 4 (2010), 58.

\bibitem[NCSG11]{navarro2011perceptual}
\textsc{Navarro F., Castillo S., Ser{\'o}n F.~J., Gutierrez D.}:
\newblock Perceptual considerations for motion blur rendering.
\newblock \emph{ACM Transactions on Applied Perception (TAP) 8}, 3 (2011),
  1--15.

\bibitem[NSG11]{navarro2011motion}
\textsc{Navarro F., Ser{\'o}n F.~J., Gutierrez D.}:
\newblock Motion blur rendering: State of the art.
\newblock In \emph{Computer Graphics Forum} (2011), vol.~30, Wiley Online
  Library, pp.~3--26.

\bibitem[O'd07]{o2007value}
\textsc{O'day D.~H.}:
\newblock The value of animations in biology teaching: a study of long-term
  memory retention.
\newblock \emph{CBE—Life Sciences Education 6}, 3 (2007), 217--223.

\bibitem[PS18]{palan2018prolific}
\textsc{Palan S., Schitter C.}:
\newblock Prolific. ac—a subject pool for online experiments.
\newblock \emph{Journal of Behavioral and Experimental Finance 17} (2018),
  22--27.

\bibitem[RBM{\etalchar{*}}20]{rekik2020decreasing}
\textsc{Rekik G., Belkhir Y., Mnif M., Masmoudi L., Jarraya M.}:
\newblock Decreasing the presentation speed of animated soccer scenes does not
  always lead to better learning outcomes in young players.
\newblock \emph{International Journal of Human--Computer Interaction 36}, 8
  (2020), 717--724.

\bibitem[RCL09]{ruiz2009computer}
\textsc{Ruiz J.~G., Cook D.~A., Levinson A.~J.}:
\newblock Computer animations in medical education: a critical literature
  review.
\newblock \emph{Medical education 43}, 9 (2009), 838--846.

\bibitem[RGR17]{rietzler2017matrix}
\textsc{Rietzler M., Geiselhart F., Rukzio E.}:
\newblock The matrix has you: realizing slow motion in full-body virtual
  reality.
\newblock In \emph{Proceedings of the 23rd ACM Symposium on Virtual Reality
  Software and Technology} (2017), pp.~1--10.

\bibitem[RHA17]{rothe2017diegetic}
\textsc{Rothe S., Hu{\ss}mann H., Allary M.}:
\newblock Diegetic cues for guiding the viewer in cinematic virtual reality.
\newblock In \emph{Proceedings of the 23rd ACM Symposium on Virtual Reality
  Software and Technology} (2017), pp.~1--2.

\bibitem[RLN07]{rosenholtz2007measuring}
\textsc{Rosenholtz R., Li Y., Nakano L.}:
\newblock Measuring visual clutter.
\newblock \emph{Journal of vision 7}, 2 (2007), 17--17.

\bibitem[Ros99]{rosenholtz1999simple}
\textsc{Rosenholtz R.}:
\newblock A simple saliency model predicts a number of motion popout phenomena.
\newblock \emph{Vision research 39}, 19 (1999), 3157--3163.

\bibitem[SBE{\etalchar{*}}14]{stengel2014temporal}
\textsc{Stengel M., Bauszat P., Eisemann M., Eisemann E., Magnor M.}:
\newblock Temporal video filtering and exposure control for perceptual motion
  blur.
\newblock \emph{IEEE Transactions on Visualization and Computer Graphics 21}, 5
  (2014), 663--671.

\bibitem[SLP{\etalchar{*}}22]{sutton2022look}
\textsc{Sutton J., Langlotz T., Plopski A., Zollmann S., Itoh Y., Regenbrecht
  H.}:
\newblock Look over there! investigating saliency modulation for visual
  guidance with augmented reality glasses.
\newblock In \emph{Proceedings of the 35th Annual ACM Symposium on User
  Interface Software and Technology} (2022), pp.~1--15.

\bibitem[SNMH13]{sharan2013simulated}
\textsc{Sharan L., Neo Z.~H., Mitchell K., Hodgins J.~K.}:
\newblock Simulated motion blur does not improve player experience in racing
  game.
\newblock In \emph{Proceedings of Motion on Games}. 2013, pp.~149--154.

\bibitem[SSS{\etalchar{*}}20]{solteszova2020memento}
\textsc{Solteszova V., Smit N.~N., Stoppel S., Gr{\"u}ner R., Bruckner S.}:
\newblock Memento: Localized time-warping for spatio-temporal selection.
\newblock In \emph{Computer Graphics Forum} (2020), vol.~39, Wiley Online
  Library, pp.~231--243.

\bibitem[ST92]{stone1992human}
\textsc{Stone L.~S., Thompson P.}:
\newblock Human speed perception is contrast dependent.
\newblock \emph{Vision research 32}, 8 (1992), 1535--1549.

\bibitem[TKL13]{tang2013teleoperated}
\textsc{Tang T., Kurkowski J., Lienkamp M.}:
\newblock Teleoperated road vehicles: A novel study on the effect of blur on
  speed perception.
\newblock \emph{International Journal of Advanced Robotic Systems 10}, 9
  (2013), 333.

\bibitem[TMB02]{tversky2002animation}
\textsc{Tversky B., Morrison J.~B., Betrancourt M.}:
\newblock Animation: can it facilitate?
\newblock \emph{International journal of human-computer studies 57}, 4 (2002),
  247--262.

\bibitem[VL15]{vivo2015book}
\textsc{Vivo P.~G., Lowe J.}:
\newblock The book of shaders.
\newblock \emph{\url{https://thebookofshaders.com}} (2015).

\bibitem[VPC08]{vaziri2008apparent}
\textsc{Vaziri-Pashkam M., Cavanagh P.}:
\newblock Apparent speed increases at low luminance.
\newblock \emph{Journal of Vision 8}, 16 (2008), 9--9.

\bibitem[WKG17]{waldner2017exploring}
\textsc{Waldner M., Karimov A., Gr{\"o}ller E.}:
\newblock Exploring visual prominence of multi-channel highlighting in
  visualizations.
\newblock In \emph{Proceedings of the 33rd Spring Conference on Computer
  Graphics} (2017), pp.~1--10.

\bibitem[WLMB{\etalchar{*}}14]{waldner2014attractive}
\textsc{Waldner M., Le~Muzic M., Bernhard M., Purgathofer W., Viola I.}:
\newblock Attractive flicker—guiding attention in dynamic narrative
  visualizations.
\newblock \emph{IEEE Transactions on Visualization and Computer Graphics 20},
  12 (2014), 2456--2465.

\bibitem[XMWZ19]{xie2019coordinating}
\textsc{Xie H., Mayer R.~E., Wang F., Zhou Z.}:
\newblock Coordinating visual and auditory cueing in multimedia learning.
\newblock \emph{Journal of Educational Psychology 111}, 2 (2019), 235.

\end{thebibliography}


\end{document}